\documentclass[aps,prb,reprint,amsmath,amssymb,floatfix]{revtex4-2}

\usepackage{graphicx}
\usepackage{dcolumn}
\usepackage{bm}
\usepackage{hyperref}
\usepackage{xcolor}

\begin{document}

\title{Fast contracted Clebsch--Gordan tensor products for equivariant graph neural networks}

\author{Anton Bochkarev}
\affiliation{ICAMS, Ruhr-Universit\"at Bochum, 44780 Bochum, Germany}

\author{Yury Lysogorskiy}
\affiliation{ICAMS, Ruhr-Universit\"at Bochum, 44780 Bochum, Germany}

\author{Ralf Drautz}
\affiliation{ICAMS, Ruhr-Universit\"at Bochum, 44780 Bochum, Germany}

\date{\today}

\begin{abstract}
We present an $\mathcal{O}(L^3)$ algorithm for evaluating contracted Clebsch--Gordan tensor products in $\mathrm{O}(3)$-equivariant machine learning potentials at fixed Canonical Polyadic (CP) rank. Mapping the angular integral to a structured Gauss--Legendre and Fourier tensor-product grid decouples the radial channel contractions from the angular transforms. The antisymmetric parity-odd Clebsch--Gordan channels, unreachable by the symmetric pointwise product on a scalar $S^2$ grid, are recovered through the surface-curl pairing $\hat r \cdot [\nabla_{S^2} A \times \nabla_{S^2} B]$, the spherical Poisson bracket, which supplies the $L=1$ angular momentum on the grid while preserving rotational equivariance. The construction extends to parity-aware equivariant message passing in architectures of the atomic cluster expansion family and is verified by direct numerical quadrature. The full uncontracted Clebsch--Gordan tensor product remains subject to the $\mathcal{O}(L^4)$ output-size lower bound. A benchmark shows wall-clock scaling empirically as $L^2$ across the practical $l_{\max}$ range. For the on-site contraction this is pre-asymptotic, giving way to $L^3$ at large $l_{\max}$. For message passing it is structural and the runtime is memory-bandwidth bound on $L^2$-sized grid tensors.
\end{abstract}

\maketitle

\section{Introduction}
The development of $\mathrm{E}(3)$-equivariant machine learning potentials, such as the Atomic Cluster Expansion (ACE)~\cite{drautz2019} and related message-passing architectures~\cite{cohenWelling2016, weiler2018, anderson2019, schuett2021, brandstetter2022, batzner2022, batatia2022, liao2023, musaelian2023, bochkarev2024, liao2024, fu2025, batatia2025, liao2026equiformerv3scalingefficientexpressive}, has transformed atomistic simulations. These models achieve high accuracy and data efficiency by preserving the physical symmetries of the underlying 3D geometries. A central operation in these architectures is the coupling of $\mathrm{O}(3)$ irreducible representations (irreps) via the Clebsch--Gordan tensor product. Implementations of the required spherical tensor transformations are available in dedicated libraries and simulation software~\cite{thomas2018, geiger2022, unke2024, lysogorskiy2021, bochkarev2022, bochkarev2024, gracemaker, lysogorskiy2026, batatia2022, witt2023, cuequivariance}.

However, the strict symmetry imposes a severe computational bottleneck. A naive implementation of the full Clebsch--Gordan tensor product scales as $\mathcal{O}(L^6)$ with the maximum angular momentum degree $L = l_{\max} + 1$. While sparse execution reduces this to $\mathcal{O}(L^5)$, angular momentum coupling remains a dominant computational cost for high-resolution models. 

Recent work has demonstrated that substituting the analytical Clebsch--Gordan contraction with Fast Fourier Transforms on a structured spherical grid~\cite{driscollHealy1994, cohen2018} can achieve $\mathcal{O}(L^3)$ scaling~\cite{passaro2023, luo2024}. While asymptotically faster Fast Spherical Harmonic Transforms scaling as $\mathcal{O}(L^2 \log^2 L)$ exist~\cite{driscollHealy1994}, they rely on recursive divide-and-conquer algorithms that perform poorly on modern GPU architectures at typical atomistic resolutions ($l_{\max} \le 20$), making dense $\mathcal{O}(L^3)$ formulations preferable for wall-clock efficiency. These $\mathcal{O}(L^3)$ methods address different aspects of equivariant machine learning potentials. The eSCN reduction~\cite{passaro2023} sparsifies the SO(3) tensor product by per-edge axis alignment, while the scalar Gaunt approach of Luo et al.~\cite{luo2024} captures the parity-even Gaunt-reachable channels and so does not reach the full set of parity-odd Clebsch--Gordan outputs.
Hot-Ham~\cite{liang2026} combines the Gaunt tensor product with local coordinate transformations for electronic-structure Hamiltonian regression, reaching $\mathcal{O}(L^3)$ to $\mathcal{O}(L^2\log^2 L)$, but inheriting the same parity-even restriction. Closed-form integral formulas for the antisymmetric Gaunt-like coefficients that recover these parity-odd channels have recently been derived~\cite{xie2026, heyraud2026}. A complementary line of work approximates the analytical Clebsch--Gordan tensor itself. Tensor decomposition networks~\cite{lin2025} replace the Clebsch--Gordan coefficient tensor by a low-rank CP factorization, reaching $\mathcal{O}(L^4)$ at the price of trading exact for approximate equivariance with controllable error. Approximate symmetry can also be targeted by data-augmentation procedures, with rigorous error analysis~\cite{klinteback2026} showing that exact symmetry preservation generally requires Haar-quadrature-style schemes whereas random-sampling augmentation only achieves square-root convergence in the symmetrization error.  
Outside the local message-passing setting, the computational bottleneck of modeling global geometric context is driving the development of sub-quadratic sequence architectures. Equivariant attention networks~\cite{fuchs2020} and long-convolutional models such as the SE(3)-Hyena operator~\cite{moskalevSE3HyenaOperatorScalable2024} are using fast Clebsch--Gordan primitives to achieve $\mathcal{O}(N\log N)$ global token complexity~\cite{howell2025}, indicating that the demand for fast Clebsch--Gordan operations is increasingly pervasive across architectural families.

In this work, we present an $\mathcal{O}(L^3)$ algorithm for evaluating contracted Clebsch--Gordan tensor products at fixed Canonical Polyadic (CP) decomposition rank. A structured tensor-product grid (Gauss--Legendre quadrature combined with Fourier sums) decouples the radial and angular transformations. The symmetric pointwise product on a scalar $\mathrm{S}^2$ grid cannot reach the antisymmetric parity-odd channels. We achieve this with a surface-curl pairing, equivalently to the spherical Poisson bracket on $S^2$. An equivalent result has been demonstrated independently by Heyraud et al.~\cite{heyraud2026}. This enables parity-aware $\mathrm{O}(3)$-equivariant operations at $\mathcal{O}(L^3)$ scaling across the broad class of CP-factorized $\mathrm{E}(3)$-equivariant machine learning potentials. Throughout, ``$\mathcal{O}(L^3)$ scaling'' refers to the leading-order cost of evaluating a contracted bilinear coupling at fixed CP rank $C$ and radial-channel count $N$. The total cost contains sub-leading $\mathcal{O}(L^2)$ contributions, which dominate the wall-clock across the practical $l_{\max}$ range, and the effective scaling is expected as $L^2$ in this regime. The contracted output has size $\mathcal{O}(L^2)$ rather than the $\mathcal{O}(L^4)$ output of the full uncontracted Clebsch--Gordan tensor product (CGTP) studied by Xie et al.~\cite{xie2026}, and the $\mathcal{O}(L^4)$ output-size lower bound established there does not constrain the contracted operation, and an $\mathcal{O}(L^3)$ algorithm is admissible. We develop this point in Sec.~\ref{sec:lower-bound-discussion}. We refer to this construction as the SCALE-CG algorithm (Surface-Curl Augmented Legendre Evaluation for Clebsch--Gordan tensor products). Initial benchmarks show wall-clock advantages over a sparse direct-sum baseline from small $L$ onwards.

The paper is structured as follows. In Sec.~\ref{sec:coupling-gnn} we introduce two coupling operations whose efficient evaluation is the subject of this paper, the on-site bilinear contraction and the parity-aware edge-to-node message-passing coupling. Sec.~\ref{sec:cg-int} reviews the relation between the Clebsch--Gordan coefficient and the Gaunt integral over three spherical harmonics, where we derive the parity selection rule and establish the surface-curl identity that recovers the parity-odd Clebsch--Gordan channels missed by the scalar spherical harmonics (SH) integral. Sec.~\ref{sec:lebedev-O4} describes an $\mathcal{O}(L^4)$ implementation on a Lebedev quadrature grid as a stepping-stone, and in Sec.~\ref{sec:structured} we replace the unstructured Lebedev grid by a structured Gauss--Legendre $\times$ Fourier tensor-product grid that achieves $\mathcal{O}(L^3)$. In Sec.~\ref{sec:msgpass-parity-cases} we derive the four parity cases of the parity-aware coupling, apply the surface-curl identity to the parity-odd channels, and treat the two-parity-bearing-fields generalization. Sec.~\ref{sec:pipelines} collects the step-by-step computational pipelines, with cost analysis. In Sec.~\ref{sec:empirical-scaling} we benchmark the algorithm numerically. Sec.~\ref{sec:summary} summarizes the results. The proof of the surface-curl identity and the numerical validation are collected in the appendices.

\section{Coupling in equivariant graph neural networks}
\label{sec:coupling-gnn}

The atomic cluster expansion~\cite{drautz2019,drautz2020,dusson2022}, NequIP~\cite{batzner2022}, multi-ACE/MACE~\cite{batatia2025, batatia2022}, GRACE~\cite{bochkarev2024}, and other equivariant graph neural networks for atomistic systems~\cite{anderson2019, schuett2021, brandstetter2022, liao2023, musaelian2023, fu2025} all build on covariant many-body features, which can be obtained by repeated application of two coupling operations. The first is the bilinear contraction of two atomic-base tensors,
\begin{multline}
\label{eq:phi-cg}
\varphi_{inlm,p} = \sum_{n_1 n_2 l_1 l_2}\,\sum_{p_1 p_2}^{p_1 p_2 = p}\, c^{(nl,p)}_{n_1 l_1 p_1\, n_2 l_2 p_2}\,\sum_{m_1 m_2} G_{l_1 m_1\, l_2 m_2}^{l m} \\
\times A^{(1)}_{i n_1 l_1 m_1, p_1}\,A^{(2)}_{i n_2 l_2 m_2, p_2}\,,
\end{multline}
in which two fields $A^{(1)}, A^{(2)}$ on the same site $i$, expanded in spherical harmonics and carrying parity labels $p_1, p_2 \in \{+1, -1\}$, are coupled to a target irrep $(l, p)$ via Clebsch--Gordan coefficients $G^{lm}_{l_1 m_1\, l_2 m_2}$. The output parity is fixed as $p = p_1 p_2$, independent of $l$ (Eq.\eqref{eq:cg-parity-rule}). For identical inputs $A^{(1)} = A^{(2)} = A$, Eq.\eqref{eq:phi-cg} produces the ACE order-2 cluster basis~\cite{drautz2019}. The learnable channel-mixing tensor $c^{(nl,p)}_{n_1 l_1 p_1\, n_2 l_2 p_2}$ specifies which input radial--angular--parity combinations contribute to each output $(n,l,p)$ slot.

The second operation is the parity-aware message-passing coupling,
\begin{multline}
\label{eq:msgpass-parity-start}
\varphi_{inlm,p} \;=\; \sum_{n_1 n_2 l_1 l_2,\,p_2}\,c^{(nl,p)}_{n_1 l_1\, n_2 l_2,\,p_2}\sum_{m_1 m_2} G^{lm}_{l_1 m_1\, l_2 m_2} \\
\times \sum_j R_{n_1 l_1}(r_{ji})\,Y_{l_1 m_1}(\hat r_{ji})\,I_{j n_2 l_2 m_2,\,p_2}\,,
\end{multline}
in which an edge feature $R_{n_1 l_1}(r_{ji})\,Y_{l_1 m_1}(\hat r_{ji})$ on the bond between central atom $i$ and neighbor $j$ (with $\vec{r}_{ji} = \vec{r}_j - \vec{r}_i$) is coupled to the neighbor's node feature $I_{j n_2 l_2 m_2, p_2}$ and pooled over the neighbors of atom~$i$. The edge feature has natural parity $(-1)^{l_1}$ and the node feature has parity $p_2$. The output parity is $p = (-1)^{l_1} p_2$.

In both Eq.\eqref{eq:phi-cg} and Eq.\eqref{eq:msgpass-parity-start} the dominant computational cost lies in the explicit Clebsch--Gordan contraction over $(l_1, l_2, m_1, m_2)$, which na\"\i vely scales as $\mathcal{O}(L^6)$ and as $\mathcal{O}(L^5)$ when the $m_1+m_2=m$ selection rule is exploited. The $\mathcal{O}(L^3)$ scaling derived in the remainder of this paper preserves exact $\mathrm{O}(3)$ equivariance and rests on (i)~re-expressing both operations as quadratures on a structured Gauss--Legendre $\times$ Fourier spherical grid, and (ii)~a CP decomposition of the channel-mixing tensor at fixed rank $C$,
\begin{equation}
\label{eq:cp-decomp}
c^{(nl,p)}_{n_1 l_1 p_1\, n_2 l_2 p_2} \;=\; \sum_c \lambda_c^{(nl,p)}\,c^{(1)}_{c n_1 l_1 p_1}\,c^{(2)}_{c n_2 l_2 p_2}\,,
\end{equation}
which is efficient in ACE-family models~\cite{drautz2019, bochkarev2022mlACE, batatia2022, darby2023, bochkarev2024, batatia2025}.

\section{Integral over Spherical Harmonics and Clebsch--Gordan Coefficients}
\label{sec:cg-int}
The integral of the product of three complex spherical harmonics over the solid angle $\Omega$ is analytically proportional to the product of two Clebsch--Gordan (CG) coefficients~\cite{varshalovich1988},
\begin{multline}
\label{eq:cg-integral}
\int Y_{l_1 m_1}(\Omega) Y_{l_2 m_2}(\Omega) Y_{lm}^*(\Omega) \, d\Omega \\
= \sqrt{\frac{(2l_1+1)(2l_2+1)}{4\pi(2l+1)}} G_{l_1 0\, l_2 0}^{l 0}\,G_{l_1 m_1\, l_2 m_2}^{l m} \,.
\end{multline}
The left-hand side of Eq.\eqref{eq:cg-integral} is the Gaunt coefficient
\begin{equation}
\label{eq:gaunt}
\mathcal{G}^{lm}_{l_1 m_1\, l_2 m_2} \;=\; \int Y_{l_1 m_1}(\Omega)\,Y_{l_2 m_2}(\Omega)\,Y_{lm}^*(\Omega)\,d\Omega\,.
\end{equation}

The first Clebsch--Gordan factor $G_{l_1 0\, l_2 0}^{l 0}$ on the right-hand side of Eq.\eqref{eq:cg-integral} is the parity-projection of the SH triple product. When $l_1 + l_2 + l$ is even, 
\begin{multline}
\label{eq:cg-explicit}
G_{l_1 0\, l_2 0}^{l 0} = (-1)^{l_1 - l_2 + g} \sqrt{2l + 1}\,\sqrt{\Delta(l_1, l_2, l)} \\
\times \frac{g!}{(g - l_1)!\,(g - l_2)!\,(g - l)!}\,,
\end{multline}
with half-sum $g = (l_1 + l_2 + l)/2$ and where the Triangle Coefficient
\begin{multline}
\label{eq:triangle}
\Delta(l_1, l_2, l) \\
= \frac{(l_1 + l_2 - l)!\,(l_1 - l_2 + l)!\,(-l_1 + l_2 + l)!}{(l_1 + l_2 + l + 1)!}
\end{multline}
is non-zero in the triangle range $|l_1 - l_2| \le l \le l_1 + l_2$.

When $l_1 + l_2 + l$ is odd, the integrand $Y_{l_1 m_1} Y_{l_2 m_2} Y_{lm}^*$ is parity-odd under inversion $\hat r \to -\hat r$, so its integral vanishes,
\begin{equation}
\label{eq:parity-zero}
G_{l_1 0\, l_2 0}^{l 0}\;=\;0 \qquad (l_1 + l_2 + l\ \text{odd})\,.
\end{equation}
Eq.\eqref{eq:cg-integral} vanishes on the parity-odd CG support, even though the underlying coefficients $G^{lm}_{l_1 m_1\, l_2 m_2}$ on that support are generically non-zero.

The parity selection rule Eq.\eqref{eq:parity-zero} is one instance of a more general feature of $\mathrm{O}(3)$ representation theory that becomes essential once fields carry explicit parity labels. The rotation--inversion group $\mathrm{O}(3) = \mathrm{SO}(3) \times \mathbb{Z}_2$ has irreps labeled by $(l, p)$ with $l \in \{0, 1, 2, \dots\}$ and parity $p \in \{+1, -1\}$, of dimension $2l+1$. The complex spherical harmonic $Y_{lm}$ satisfies $Y_{lm}(-\hat r) = (-1)^l Y_{lm}(\hat r)$ and so realizes the natural-parity irrep $(l, (-1)^l)$. The opposite-parity irreps $(l, -(-1)^l)$, namely the pseudoscalar at $l=0$, the axial vector at $l=1$, and so on, are not representable as scalar functions on $S^2$ at fixed degree. Reaching them by quadrature requires going beyond pointwise products of scalar fields.

For a CG coupling $(l_1, p_1) \otimes (l_2, p_2) \to (l, p)$ in $\mathrm{O}(3)$, the output parity is
\begin{equation}
\label{eq:cg-parity-rule}
p = p_1\,p_2\,,
\end{equation}
independent of $l$. For natural-parity scalar-SH inputs ($p_i = (-1)^{l_i}$), the output is at natural parity $(-1)^l$ when $l_1 + l_2 + l$ is even and at opposite parity when odd. The latter is the regime where Eq.\eqref{eq:cg-integral} vanishes.

This carries a direct consequence for the bilinear contraction Eq.\eqref{eq:phi-cg} and the message-passing coupling Eq.\eqref{eq:msgpass-parity-start}. For the message-passing case, $(l_1, (-1)^{l_1}) \otimes (l_2, p_2) \to (l, p)$ has $p = (-1)^{l_1} p_2$, resulting in four cases summarized in Table~\ref{tab:parity-cases-overview}.
\begin{table}[h]
\centering
\begin{tabular}{c c c c c}
\hline
$p_2$ & $l_1+l_2+l$ & $p$ &    & Gaunt-reachable? \\
\hline
$(-1)^{l_2}$  & even & $(-1)^l$    & natural  & yes \\
$(-1)^{l_2}$  & odd  & $-(-1)^l$   & opposite & no \\
$-(-1)^{l_2}$ & even & $-(-1)^l$   & opposite & yes \\
$-(-1)^{l_2}$ & odd  & $(-1)^l$    & natural  & no \\
\hline
\end{tabular}
\caption{The four parity cases of the message-passing coupling Eq.\eqref{eq:msgpass-parity-start}. Only the rows with $l_1+l_2+l$ even are reachable by Eq.\eqref{eq:cg-integral}.}
\label{tab:parity-cases-overview}
\end{table}
Each output parity sector receives one Gaunt-reachable and one parity-odd contribution, the latter requiring an alternative grid construction.

\subsection*{Surface-curl identity for the parity-odd channels}
\label{sec:parity-odd-identity}

The CG coefficients in the parity-odd channel satisfy
\begin{multline}
\label{eq:CG-antisym}
G^{lm}_{l_1m_1\,l_2m_2} = (-1)^{l_1+l_2-l}\,G^{lm}_{l_2m_2\,l_1m_1} = -G^{lm}_{l_2m_2\,l_1m_1} \\
(l_1+l_2+l\text{ odd})\,,
\end{multline}
i.e.\ they are antisymmetric under exchange of the two input slots. The canonical example is the $1\otimes 1\to 1$ cross product $G^{1m}_{1m_1\,1m_2} \propto \epsilon_{m_1 m_2 m}$. In contrast, any bilinear scalar product $A^{(1)}(\hat r) A^{(2)}(\hat r)$ on the grid is symmetric under $1\leftrightarrow 2$. We resolve this by surface-curl coupling, in which the two factors are differentiated tangentially and the gradients are crossed against $\hat r$. In coordinates $(\theta, \phi)$ the radial-projected cross product collapses to a $2\times 2$ Jacobian, the spherical Poisson bracket of $A^{(1)}$ and $A^{(2)}$,
\begin{multline}
\label{eq:poisson-S2}
\{A^{(1)}, A^{(2)}\}_{S^2} \;\equiv\; \hat r \cdot \bigl[\nabla_{S^2} A^{(1)} \times \nabla_{S^2} A^{(2)}\bigr] \\
\;=\; \frac{1}{\sin\theta}\bigl(\partial_\theta A^{(1)}\,\partial_\phi A^{(2)} \;-\; \partial_\phi A^{(1)}\,\partial_\theta A^{(2)}\bigr)\,,
\end{multline}
where $\nabla_{S^2}$ is the surface gradient on the unit sphere. With $x = \cos\theta$ and $\partial_\theta = -\sin\theta\,\partial_x$, the bracket becomes
\begin{equation}
\label{eq:poisson-S2-x}
\{A^{(1)}, A^{(2)}\}_{S^2} \;=\; \partial_\phi A^{(1)}\,\partial_x A^{(2)} \;-\; \partial_x A^{(1)}\,\partial_\phi A^{(2)}\,,
\end{equation}
free of the $1/\sin\theta$ factor and natively suited to the Gauss--Legendre grid in $x$. Under $1\leftrightarrow 2$ the bracket changes sign, matching the antisymmetry of the parity-odd CG, and as the symplectic structure on $S^2$ realized as a coadjoint orbit of $\mathrm{SO}(3)$~\cite{varshalovich1988} it is rotation-equivariant. Substituting Eq.\eqref{eq:poisson-S2} into the SH integral, we consider at given $(l_1, l_2, l)$ on the parity-odd support ($l_1+l_2+l$ odd) the integral
\begin{multline}
\label{eq:Q-def}
Q^{lm}_{l_1 m_1\, l_2 m_2} = \\ \int \hat r \cdot \bigl[\nabla_{S^2} Y_{l_1 m_1}(\hat r) \times \nabla_{S^2} Y_{l_2 m_2}(\hat r)\bigr] Y_{lm}^*(\hat r)\,d\Omega \,.
\end{multline}
This is an SO(3)-equivariant tensor with magnetic indices $(m, m_1, m_2)$, transforming as the coupling $D^{l_1} \otimes D^{l_2} \to D^l$ and antisymmetric under $(l_1 m_1) \leftrightarrow (l_2 m_2)$. Since the $D^l$ summand occurs with multiplicity one in $D^{l_1} \otimes D^{l_2}$, the Wigner--Eckart theorem forces $Q^{lm}_{l_1 m_1\, l_2 m_2}$ to be proportional to the unique covariant tensor with these properties at fixed $(l_1, l_2, l)$, the Clebsch--Gordan coefficient Eq.\eqref{eq:CG-antisym},
\begin{equation}
\label{eq:Q-CG}
Q^{lm}_{l_1 m_1\, l_2 m_2} \;=\; \kappa^{-1}(l_1, l_2, l)\,G^{lm}_{l_1 m_1\, l_2 m_2}\,,
\end{equation}
where $\kappa^{-1}(l_1, l_2, l)$ depends only on the SH and CG conventions. Therefore, from Eq.\eqref{eq:Q-CG},
\begin{multline}
\label{eq:parity-odd-identity}
G^{lm}_{l_1m_1\,l_2m_2} = \kappa(l_1, l_2, l) \int d\Omega\, Y_{lm}^*(\hat r) \\
\times \hat r \cdot \bigl[\nabla_{S^2}Y_{l_1m_1}(\hat r) \times \nabla_{S^2}Y_{l_2m_2}(\hat r)\bigr]\,,
\end{multline}
valid on the parity-odd support. A proof using $\mathrm{SO}(3)$ covariance, Schur's lemma, and inversion parity is given in Appendix~\ref{sec:appendix-derivation}. An analogous identity has been obtained by Heyraud et al.~\cite{heyraud2026} via simplification of the vector spherical tensor product. We further confirm Eq.\eqref{eq:parity-odd-identity} by direct numerical quadrature in Appendix~\ref{sec:numerical-validation}, where every parity-odd triple yields a unique $\kappa^{-1}(l_1, l_2, l)$ that is purely imaginary in the complex SH basis, and the right-hand side reproduces $G^{lm}_{l_1m_1\,l_2m_2}$ to numerical precision over all $(m_1, m_2, m)$. The magnitude $|\kappa^{-1}|$ is unitary-invariant. In a real SH basis $\kappa^{-1}$ is real and the entire transformation runs in real arithmetic (Appendix~\ref{sec:realsh}).

The derivative fields $\partial_\phi A$ and $\partial_x A$ are obtained by applying $\partial_x \tilde P_l^{|m|}(x)$ and the rule $\partial_\phi \to im$ to the separated SH expansion. The surface-curl mechanism connects to the vector-spherical-harmonic constructions of Xie et al.~\cite{xie2026}, where the parity-odd content lives in the tangent-vector structure of the input fields rather than in their scalar amplitudes.

\section{Representation using Lebedev Weighted Sum - $\mathcal{O}(L^4)$}
\label{sec:lebedev-O4}
This section and the next develop the parity-even ($l_1+l_2+l$ even), Gaunt-reachable angular momenta of Eq.\eqref{eq:phi-cg} with both inputs in their natural parity. The parity-odd channels and the general parity-labeled case are recovered by the surface-curl identity Eq.\eqref{eq:parity-odd-identity} and treated together with the message-passing operation Eq.\eqref{eq:msgpass-parity-start} in Sec.~\ref{sec:msgpass-parity-cases}. 

To compute the geometric integral Eq.\eqref{eq:cg-integral}, we replace the continuous integral with a discrete sum over points on the sphere. We choose a Lebedev grid of $K$ points $\Omega_k$ with weights $w_k$ \cite{lebedev1976},
\begin{multline}
\label{eq:lebedev-sum}
\int Y_{l_1 m_1} Y_{l_2 m_2} Y_{lm}^* \, d\Omega \\
= \sum_{k=1}^K w_k\, Y_{l_1 m_1}(\Omega_k)\, Y_{l_2 m_2}(\Omega_k)\, Y_{lm}^*(\Omega_k) \,.
\end{multline}
This discrete summation is exact provided the chosen Lebedev grid has a degree of exactness $D \ge l_1 + l_2 + l$, which implies that $K$ is $\mathcal{O}(L^2)$ \cite{lebedev1976}. 

We define $y_{klm}$ as the value of spherical harmonics at Lebedev points
\begin{equation}
\label{eq:yklm}
y_{klm} = Y_{l m}(\Omega_k) \,.
\end{equation}
Substituting Eq.\eqref{eq:cg-integral} and Eq.\eqref{eq:lebedev-sum} into Eq.\eqref{eq:phi-cg} gives
\begin{multline}
\label{eq:phi-with-f}
\varphi_{inlm} = \sum_{n_1 n_2 l_1 l_2} c^{(nl)}_{n_1 l_1\, n_2 l_2}\,f(l_1, l_2, l) \\
\times \sum_{m_1 m_2} \sum_k w_k\, y_{k l_1 m_1}\, y_{k l_2 m_2}\, y_{klm}^* \\
\times A^{(1)}_{i n_1 l_1 m_1}\,A^{(2)}_{i n_2 l_2 m_2}\,,
\end{multline}
where $f(l_1, l_2, l)$ is the prefactor relating the spherical-harmonic integral to the Clebsch--Gordan coefficient via Eq.\eqref{eq:cg-integral},
\begin{equation}
\label{eq:f-def}
f(l_1, l_2, l) \;=\; \sqrt{\frac{4\pi(2l+1)}{(2l_1+1)(2l_2+1)}}\,\frac{1}{G^{l 0}_{l_1 0\, l_2 0}}\,,
\end{equation}
defined to be zero on the parity-forbidden support where $G^{l 0}_{l_1 0\, l_2 0}=0$.
Applying a CP decomposition at rank $C$ to the prefactor-weighted coupling tensor $c^{(nl)}_{n_1l_1\,n_2l_2}\,f(l_1,l_2,l)$, which is parameterized directly in this form, results in
\begin{multline}
\label{eq:phi-substituted}
\varphi_{inlm} = \sum_k w_k\, y_{klm}^*  \sum_c \lambda_c^{(nl)} \\
\times \sum_{n_1 l_1 m_1} c^{(1)}_{c n_1 l_1}\, y_{k l_1 m_1}\, A^{(1)}_{i n_1 l_1 m_1} \\
\times \sum_{n_2 l_2 m_2} c^{(2)}_{c n_2 l_2}\, y_{k l_2 m_2}\, A^{(2)}_{i n_2 l_2 m_2}\,.
\end{multline}
This eliminates the contraction with the Clebsch--Gordan coefficients. The corresponding pipeline and scaling analysis is given in Sec.~\ref{sec:pipeline-lebedev}.

\section{Representation using Structured Tensor-Product Grids - $\mathcal{O}(L^3)$}
\label{sec:structured}

To reduce the complexity from $\mathcal{O}(L^4)$ to $\mathcal{O}(L^3)$, we replace the unstructured Lebedev grid with a structured tensor-product grid that allows us to exploit the separation of variables in the spherical harmonics,
\begin{equation}
\label{eq:Ylm-separated}
Y_{lm}(\theta, \phi) = \tilde{P}_l^{|m|}(\cos\theta)\, e^{im\phi} \,.
\end{equation}

We define a grid composed of $U$ Gauss--Legendre quadrature points for the polar angle $\theta_u$ (with corresponding weights $w_u$) and $V$ uniformly spaced points for the azimuthal angle $\phi_v = 2\pi v / V$. To ensure exact integration, we require $U \ge (3 l_{\max} + 1)/2$ and $V \ge 3 l_{\max} + 1$ \cite{driscollHealy1994}. The total number of grid points is $K = U \times V = \mathcal{O}(L^2)$.

In the following we continue with the parity-even angular momenta of Sec.~\ref{sec:lebedev-O4}, before we address the parity-odd and general parity-labeled cases in Sec.~\ref{sec:msgpass-parity-cases}. 

Instead of a single index $k$ for the Lebedev grid, we use the separated angular components
\begin{equation}
\label{eq:pulm-qvm}
p_{ulm} = \tilde{P}_l^{|m|}(\cos\theta_u), \quad q_{vm} = e^{im\phi_v} \,,
\end{equation}
where $\tilde{P}_l^{|m|}$ are the normalized associated Legendre polynomials whose continuous inner product evaluates to $\delta_{ll'}$. Replacing the continuous Clebsch--Gordan integral with the discrete Gauss--Legendre and Fourier sums and applying a rank-$C$ CP decomposition to the prefactor-weighted coupling tensor $c^{(nl)}_{n_1l_1\,n_2l_2}\,f(l_1, l_2, l)$, the contracted basis can be rearranged as
\begin{multline}
\label{eq:struct-phi-explicit}
\varphi_{inlm} = \sum_u w_u\, p_{ulm} \sum_v \frac{2\pi}{V}\, q_{vm}^* \sum_c \lambda_c^{(nl)} \\
\times \sum_{n_1 l_1 m_1} c^{(1)}_{c n_1 l_1}\, p_{u l_1 m_1}\, q_{v m_1}\, A^{(1)}_{i n_1 l_1 m_1} \\
\times \sum_{n_2 l_2 m_2} c^{(2)}_{c n_2 l_2}\, p_{u l_2 m_2}\, q_{v m_2}\, A^{(2)}_{i n_2 l_2 m_2} \,.
\end{multline}
To eliminate the $\mathcal{O}(L^4)$ contraction, we explicitly separate the sums over $l$ and $m$. In the corresponding pipeline (Sec.~\ref{sec:pipeline-structured}), this factorization reduces the scaling to $\mathcal{O}(L^3)$. A related spherical-grid evaluation has also been used by Xie et al.~\cite{xie2025} to simplify the Gaunt tensor product of Luo et al.~\cite{luo2024}.

\section{Parity-aware grid evaluation}
\label{sec:msgpass-parity-cases}

In Secs.~\ref{sec:lebedev-O4} and~\ref{sec:structured} we developed the structured-grid evaluation of Eq.\eqref{eq:phi-cg} under the natural-parity, Gaunt-reachable angular momenta. We now treat the general case in which the inputs carry arbitrary parity labels. The four cases take Eq.\eqref{eq:msgpass-parity-start} as the example, classifying the cases by the sign of $p_2$ relative to $(-1)^{l_2}$ and the parity of $l_1+l_2+l$. The same logic applies to the on-site Eq.\eqref{eq:phi-cg} after substituting $A^{(2)}$ for $I$ and the generalization to two free parity labels $(p_1, p_2)$ in Sec.~\ref{sec:two-parity}. Each case can either be evaluated by the Gaunt structured-grid (Category $\alpha$) or the surface-curl construction Eq.\eqref{eq:parity-odd-identity} (Category $\beta$), with explicit pipelines given in Sec.~\ref{sec:pipelines}.

\subsection{Case A: natural input parity, $l_1+l_2+l$ even}

When $p_2 = (-1)^{l_2}$ and $l_1+l_2+l$ is even, the output parity is
\begin{equation}
p = (-1)^{l_1}\,(-1)^{l_2} = (-1)^l\,,
\end{equation}
i.e.\ the natural-parity sector at $l$. Both inputs are scalar spherical harmonics in their natural parity and the Gaunt integral
\begin{equation}
\int Y_{l_1m_1}\,Y_{l_2m_2}\,Y_{lm}^*\,d\Omega \;\propto\; G^{l 0}_{l_1 0\,l_2 0}\,G^{lm}_{l_1m_1\,l_2m_2} \,,
\end{equation}
is non-vanishing. This is the case handled by the structured-grid contraction of Sec.~\ref{sec:structured}. The corresponding pipelines are given in Secs.~\ref{sec:pipeline-structured} and~\ref{sec:pipeline-msgpass}.

\subsection{Case B: natural input parity, $l_1+l_2+l$ odd}

When $p_2 = (-1)^{l_2}$ and $l_1+l_2+l$ is odd, the output parity is
\begin{equation}
p = (-1)^{l_1}\,(-1)^{l_2} = -(-1)^l\,,
\end{equation}
the opposite-parity sector at $l$. The Gaunt integral vanishes and we employ the surface-curl identity Eq.\eqref{eq:parity-odd-identity}.

When $A^{(1)} = A^{(2)} = A$, the parity-odd channel does not survive the radial sum.
\begin{multline}
\label{eq:self-coupled-vanish}
\Phi^{(\text{odd})}_{nlm} \;=\; \sum_{n_1 l_1\, n_2 l_2}^{l_1 + l_2 + l\text{ odd}}\,c^{(nl)}_{n_1l_1\,n_2l_2} \\
\times \sum_{m_1 m_2} G^{lm}_{l_1m_1\,l_2m_2}\,A_{n_1 l_1 m_1}\,A_{n_2 l_2 m_2}\,,
\end{multline}
the antisymmetry of $G$ under $(l_1m_1)\leftrightarrow(l_2m_2)$ contracts against the symmetric sum over identical dummy indices $(n_1, l_1, m_1)$ and $(n_2, l_2, m_2)$, giving zero identically whenever $c^{(nl)}_{n_1l_1\,n_2l_2}$ is symmetric under $(n_1 l_1)\leftrightarrow(n_2 l_2)$.

\subsection{Case C: opposite input parity, $l_1+l_2+l$ even}

When $p_2 = -(-1)^{l_2}$ and $l_1+l_2+l$ is even, $p = -(-1)^l$ (opposite-parity sector). The input $I$ at the parity-flipped slot is a pseudo-tensor, but its $(2l_2+1)$ coefficients still rotate under $D^{l_2}$. The Gaunt integral from Case~A applies with the parity-flipped coefficient array. The opposite output parity is enforced by the slot rule, with coefficients read from $I_{jn_2l_2m_2,-(-1)^{l_2}}$ and written to $\varphi_{inlm,-(-1)^l}$.

\subsection{Case D: opposite input parity, $l_1+l_2+l$ odd}

When $p_2 = -(-1)^{l_2}$ and $l_1+l_2+l$ is odd, $p = (-1)^l$ (natural-parity sector). As in Case~B, the Gaunt integral vanishes and the surface-curl construction Eq.\eqref{eq:parity-odd-identity} applies. The pipeline is identical to Case~B but reads $I$ from the parity-flipped slot, with results routed to the natural-parity output. The construction vanishes for identical inputs with symmetric channel mixing.

\subsection{Two parity-bearing factors}
\label{sec:two-parity}

Cases A--D above fixed the first factor at natural parity $p_1 = (-1)^{l_1}$. The same construction applies to two further scenarios in which both factors carry independent parity labels: (i) the on-site contraction Eq.\eqref{eq:phi-cg} with general $(p_1, p_2)$, and (ii) the deeper message-passing variant in which the edge-side message has accumulated parity content from previous layers and no longer factors as $R\cdot Y$. The latter is written with an edge field $I^{(E)}_{jin_1l_1m_1,p_1}$ and a node field $I^{(N)}_{jn_2l_2m_2,p_2}$,
\begin{multline}
\label{eq:twoparity-phi}
\varphi_{inlm,p} = \sum_{n_1n_2l_1l_2,\,p_1,p_2}\,c^{(nl,p)}_{n_1l_1\,n_2l_2,\,p_1,p_2} \\
\times \sum_{m_1m_2}G^{lm}_{l_1m_1\,l_2m_2}\sum_j I^{(E)}_{jin_1l_1m_1,\,p_1}\,I^{(N)}_{jn_2l_2m_2,\,p_2} \,.
\end{multline}
Eq.\eqref{eq:phi-cg} is the on-site specialization, replacing $\sum_j I^{(E)} I^{(N)}$ by $A^{(1)} A^{(2)}$. The output parity is $p = p_1\,p_2$. For fixed $(l, p)$, two $(p_1, p_2)$ pairs satisfy $p_1 p_2 = p$, namely $(+1, p)$ and $(-1, -p)$. Combined with $l_1+l_2+l$ even or odd, this gives eight cases in total, splitting into Category $\alpha$ (direct Gaunt) and Category $\beta$ (surface-curl). Pipeline modifications are stated in Sec.~\ref{sec:pipeline-msgpass}.

\subsection{Relation to other algorithms}
\label{sec:lower-bound-discussion}

Xie et al.~\cite{xie2026} prove that the full Clebsch--Gordan tensor product (CGTP) is bounded below by $\mathcal{O}(L^4)$ and present an algorithm at $\mathcal{O}(L^4 \log^2 L)$ via tensor spherical harmonics. The bound applies to a specific operation. ``Full CGTP'' produces an explicit output indexed by $(l_1, l_2, l, m)$, requiring $\mathcal{O}(L^4)$ output entries, with independent learnable coefficients at every $(l_1, l_2, l)$ triple.

Eqs.~\eqref{eq:phi-cg} and~\eqref{eq:msgpass-parity-start} are a different operation. They sum over $(l_1, l_2)$ with a learnable channel-mixing tensor $c^{(nl,p)}$, producing an output indexed only by $(i, n, l, m, p)$ of size $\mathcal{O}(I_{\text{nodes}} N L^2)$. The CP decomposition Eq.\eqref{eq:cp-decomp} runs this contraction at constant CP rank $C$ independent of $L$. The $L^2$ ratio between the two output sizes accounts for the gap between $\mathcal{O}(L^3)$ here and $\mathcal{O}(L^4 \log^2 L)$ in Xie et al., modulo the $\log^2 L$ factor of their FFT-based construction.

A complementary CP-based approach to the uncontracted CGTP is the tensor decomposition network (TDN) of Lin et al.~\cite{lin2025}, which approximates the Clebsch--Gordan coefficient tensor itself by a rank-$R$ CP factorization and reaches $\mathcal{O}(L^4)$ effective scaling at the cost of approximate equivariance. SCALE-CG places its CP decomposition elsewhere. The CG coefficients are evaluated exactly through structured-grid quadrature, and the low-rank structure is exposed in the learnable channel-mixing tensor of Eq.\eqref{eq:cp-decomp} rather than in $G$ itself. The model class with CP-factorized channel mixing is an efficient design choice in the ACE-family~\cite{drautz2019, bochkarev2022mlACE, batatia2022, darby2023, bochkarev2024, batatia2025}, and materializing the full $\mathcal{O}(L^4)$ uncontracted CGTP output would be mathematically redundant before the rank-$C$ projection that immediately follows. Within this class, SCALE-CG runs at $\mathcal{O}(L^3)$ with exact equivariance for both parity-even (Category $\alpha$) and parity-odd (Category $\beta$) channels.

The generalization to the full Clebsch--Gordan content, including the parity-odd channels via the surface-curl construction Eq.\eqref{eq:parity-odd-identity}, distinguishes SCALE-CG from the scalar Gaunt-based approach of Luo et al.~\cite{luo2024}, which captures only the parity-even Gaunt-reachable channels at $\mathcal{O}(L^3)$. Heyraud et al.~\cite{heyraud2026} recently derived closed-form integral formulas for the antisymmetric analogues of the Gaunt coefficients that simplify the vector-spherical-tensor-product evaluation. A systematic analysis of the expressivity-versus-runtime tradeoffs across the Clebsch--Gordan, Gaunt, and eSCN tensor products, including a spherical-grid simplification of the Gaunt construction, is given by Xie et al.~\cite{xie2025}.

The eSCN approach of Passaro and Zitnick~\cite{passaro2023} also achieves $\mathcal{O}(L^3)$ scaling for equivariant message passing, but by a fundamentally different mechanism. eSCN aligns the source-feature primary axis with each edge direction via a Wigner-D rotation, sparsifying the Clebsch--Gordan tensor product to a sequence of $2\times 2$ SO(2) convolutions. SCALE-CG, by contrast, replaces the analytical Clebsch--Gordan contraction with quadrature on a structured spherical grid, evaluating both edge and node features on the grid and merging them pointwise.

The two constructions emphasize different aspects of equivariant transformations. eSCN offers $C\times C$ SO(2) convolutions per edge per $m$-block, providing flexible per-edge channel-mixing capacity. SCALE-CG amortizes the node-side projections once per node across all incident edges and so preserves the ACE-style density trick~\cite{drautz2019}, where the per-node SH expansion is summed over neighbors before any further coupling. The dominant cost per atom therefore does not scale with the neighbor count $Z$ on the node-side passes. 

The grid pointwise product extends to $\nu$-body features through additional pointwise multiplications on the grid followed by a single inverse transform, which fits the structure of the ACE-family. Exactness of the bilinear product requires grid bandwidth $U \gtrsim (3l_{\max}+1)/2$ and $V \gtrsim 3l_{\max}+1$. Higher-body-order products tighten this requirement linearly in the body order $\nu$ and a quantitative comparison with eSCN at production body orders is left to future work.

\section{Optimized computational pipelines}
\label{sec:pipelines}

We collect the explicit step-by-step pipelines that implement the SCALE-CG algorithm developed in the previous sections. Numerical examples and verification tests for the pipelines defined here are given in Sec.~\ref{sec:empirical-scaling} and Appendix~\ref{sec:numerical-validation}. Three pipelines are presented:
\begin{itemize}
\item Sec.~\ref{sec:pipeline-lebedev}. Bilinear contraction Eq.\eqref{eq:phi-cg} on a Lebedev grid, restricted to natural-parity inputs and the Gaunt-reachable parity-even support, $\mathcal{O}(INL^4 + ICNL^3)$.
\item Sec.~\ref{sec:pipeline-structured}. Bilinear contraction Eq.\eqref{eq:phi-cg} on a structured Gauss--Legendre $\times$ Fourier grid, again for natural-parity inputs and the Gaunt-reachable parity-even support, $\mathcal{O}(ICL^3 + ICNL^2)$.
\item Sec.~\ref{sec:pipeline-msgpass}. Parity-aware grid pipeline for the message-passing coupling Eq.\eqref{eq:msgpass-parity-start}, exploiting the factorization $E = R(r_{ji})\,Y_{l_1m_1}(\hat r_{ji})$ of the edge feature, $\mathcal{O}(I_{\rm nodes}[ZCL^3 + NCL^2])$.
\end{itemize}
The first two pipelines are the parity-even specializations of Eq.\eqref{eq:phi-cg}. For self-coupled inputs with symmetric channel-mixing the Category-$\beta$ contribution vanishes by Eq.\eqref{eq:self-coupled-vanish} and these pipelines compute the entire output. The third pipeline provides full parity coverage by combining a Category-$\alpha$ run with a Category-$\beta$ run, exploiting the $R\cdot Y$ factorization of the edge feature via $\tilde R$ rescaling to save a factor of $L$ in edge-radial work. The on-site Eq.\eqref{eq:phi-cg} with general parity labels and the two-parity-bearing-fields coupling Eq.\eqref{eq:twoparity-phi} lack this factorization on either factor. Their pipelines run two structured-grid forward passes augmented with the parity-dispatch and surface-curl machinery of Sec.~\ref{sec:pipeline-msgpass}, with modifications stated there.

\subsection{Pipeline for the bilinear contraction Eq.\eqref{eq:phi-cg} on a Lebedev grid}
\label{sec:pipeline-lebedev}

Define the grid weights
\begin{equation}
\label{eq:tildew}
\tilde w_{klm} = w_k\,y^*_{klm},
\end{equation}
precomputable once from the Lebedev grid. The forward pass then consists
of four contractions plus a pointwise product, each implementable as a
single dense GEMM call.

\textbf{Step 1: Grid evaluation.} Project each atomic basis onto the Lebedev grid and contract the radial/angular axes with the learnable CP factors $c^{(1)}_{cnl}$ and $c^{(2)}_{cnl}$,
\begin{equation}
\label{eq:lebedev-X}
X^{(s)}_{iknl} = \sum_{m=-l}^{l} y_{klm}\,A^{(s)}_{inlm}\,, \qquad s\in\{1,2\}\,,
\end{equation}
\begin{equation}
\label{eq:lebedev-tildeA}
\tilde A^{(s)}_{ick} = \sum_{nl} c^{(s)}_{cnl}\,X^{(s)}_{iknl}\,, \qquad s\in\{1,2\}\,.
\end{equation}

\textbf{Step 2: Pointwise product.}
\begin{equation}
\label{eq:lebedev-S}
S_{ick} = \tilde A^{(1)}_{ick}\,\tilde A^{(2)}_{ick}\,.
\end{equation}

\textbf{Step 3: $\lambda$-projection.} Reintroduce the output channel index via the learnable weights $\lambda^{(nl)}_c$,
\begin{equation}
\label{eq:lebedev-V}
V_{iknl} = \sum_c \lambda^{(nl)}_c\,S_{ick}\,.
\end{equation}

\textbf{Step 4: Final target projection.} Integrate against the target spherical harmonics on the grid using the precomputed weights,
\begin{equation}
\label{eq:lebedev-phi}
\phi_{inlm} = \sum_k \tilde w_{klm}\,V_{iknl} \qquad (|m|\le l)\,.
\end{equation}

\textbf{Computational scaling.} Let $I$ be the number of environments, $K$ the number of Lebedev points, $N$ the radial channels, $L = l_{\max}+1$, $C$ the CP rank. For a Lebedev grid of exactness $\ge 3 l_{\max}$, $K = \mathcal{O}(L^2)$. The cost of each step is summarized in Table~\ref{tab:lebedev-scaling}.

\begin{table}[!htbp]
\centering
\begin{tabular}{c l c}
\hline
Step & Operation & Cost \\
\hline
1a & $X^{(s)}_{iknl}=\sum_{m} y_{klm}\,A^{(s)}_{inlm}$, $s=1,2$ & $2\,I K N L^2$ \\
1b & $\tilde A^{(s)}_{ick}=\sum_{nl} c^{(s)}_{cnl}\,X^{(s)}_{iknl}$, $s=1,2$ & $2\,I K C N L$ \\
2  & $S_{ick}=\tilde A^{(1)}_{ick}\,\tilde A^{(2)}_{ick}$ & $I C K$ \\
3  & $V_{iknl}=\sum_c \lambda^{(nl)}_c\,S_{ick}$ & $I K C N L$ \\
4  & $\phi_{inlm}=\sum_k \tilde w_{klm}\,V_{iknl}$ & $I K N L^2$ \\
\hline
\end{tabular}
\caption{Computational scaling of the Lebedev pipeline.}
\label{tab:lebedev-scaling}
\end{table}

The total cost is $\mathcal{O}(IKL[N(L+1) + CN])$, which with $K=\mathcal{O}(L^2)$ becomes $\mathcal{O}(INL^4 + ICNL^3)$. Compared with the direct sparse Clebsch--Gordan implementation [$\mathcal{O}(I C N^2 L^5)$ with the $m_1+m_2=m$ rule, $\mathcal{O}(I C N^2 L^6)$ without], the asymptotic speed-up is $\mathcal{O}(C N L^2/(L+C))$, quadratic in $L$ when $L\sim C$ and linear when $C\ll L$.

\subsection{Pipeline for the bilinear contraction Eq.\eqref{eq:phi-cg} on a structured tensor-product grid}
\label{sec:pipeline-structured}

Following the separated $(l, m)$ summations of Sec.~\ref{sec:structured}, the forward pass is broken into a sequence of transforms. The complexity of any single step is at most $\mathcal{O}(L^3)$.

\textbf{Step 1: Grid evaluation (Separated Transforms).} For each input field $s\in\{1,2\}$ we contract the radial axes with the CP factor $c^{(s)}_{cnl}$, then apply the Legendre transform over $\theta$, followed by the Fourier transform over $\phi$,
\begin{equation}
\label{eq:struct-H}
H^{(s)}_{iclm} = \sum_{n} c^{(s)}_{cnl}\, A^{(s)}_{inlm} \qquad (|m|\le l)\,,
\end{equation}
\begin{equation}
\label{eq:struct-F}
F^{(s)}_{icmu} = \sum_{l=|m|}^{l_{\max}} p_{ulm}\, H^{(s)}_{iclm}\,,
\end{equation}
\begin{equation}
\label{eq:struct-tildeA}
\tilde A^{(s)}_{icuv} = \sum_{m=-l_{\max}}^{l_{\max}} q_{vm}\, F^{(s)}_{icmu}\,.
\end{equation}

\textbf{Step 2: Pointwise product.}
\begin{equation}
\label{eq:struct-S}
S_{icuv} = \tilde A^{(1)}_{icuv}\, \tilde A^{(2)}_{icuv}\,.
\end{equation}

\textbf{Step 3: Inverse Transforms.} Project the pointwise-product field back to spherical-harmonic coefficients in CP-rank space using the inverse Fourier transform over $\phi$ followed by Gauss--Legendre quadrature over $\theta$,
\begin{equation}
\label{eq:struct-Tmu}
T_{icmu} = \frac{2\pi}{V} \sum_{v=1}^{V} q_{vm}^* S_{icuv}\,,
\end{equation}
\begin{equation}
\label{eq:struct-Tlm}
T_{iclm} = \sum_{u=1}^{U} w_u p_{ulm} T_{icmu} \qquad (|m|\le l)\,.
\end{equation}

\textbf{Step 4: $\lambda$-projection.} Reintroduce the output channel index via the weights $\lambda_c^{(nl)}$,
\begin{equation}
\label{eq:struct-phi}
\varphi_{inlm} = \sum_{c} \lambda_c^{(nl)} T_{iclm} \qquad (|m|\le l)\,.
\end{equation}
Applying $\lambda_c^{(nl)}$ after the inverse transforms means they run $C$ times rather than $N$ times, and the heavy $\lambda$-contraction acts on the $\mathcal{O}(L^2)$ tensor $T_{iclm}$ rather than the $\mathcal{O}(L^3)$ grid tensor. This mirrors Step~1, where the radial CP factors are applied before the forward transforms for the same reason.

\textbf{Computational scaling.} Let $I$ be the number of environments, $N_{\rm in}$ and $N_{\rm out}$ the input and output radial-channel counts, $L = l_{\max} + 1$, and $C$ the CP rank. The structured grid has $U = \mathcal{O}(L)$ and $V = \mathcal{O}(L)$. The cost of each step is summarized in Table~\ref{tab:structured-scaling}.

\begin{table}[!htbp]
\centering
\begin{tabular}{c l c}
\hline
Step & Operation & Cost \\
\hline
1a & $H^{(s)}_{iclm} = \sum_{n} c^{(s)}_{cnl}\,A^{(s)}_{inlm}$, $s=1,2$ & $2\,IC N_{\rm in} L^2$ \\
1b & $F^{(s)}_{icmu} = \sum_{l} p_{ulm}\,H^{(s)}_{iclm}$, $s=1,2$ & $2\,ICL^3$ \\
1c & $\tilde A^{(s)}_{icuv} = \sum_{m} q_{vm}\,F^{(s)}_{icmu}$, $s=1,2$ & $2\,ICL^2 \log L$ \\
2 & $S_{icuv} = \tilde A^{(1)}_{icuv}\,\tilde A^{(2)}_{icuv}$ & $ICL^2$ \\
3a & $T_{icmu} = \frac{2\pi}{V}\sum_{v} q_{vm}^*\,S_{icuv}$ & $ICL^2 \log L$ \\
3b & $T_{iclm} = \sum_{u} w_u\,p_{ulm}\,T_{icmu}$ & $ICL^3$ \\
4 & $\varphi_{inlm} = \sum_{c} \lambda_c^{(nl)}\,T_{iclm}$ & $IC N_{\rm out} L^2$ \\
\hline
\end{tabular}
\caption{Computational scaling of the structured-grid pipeline. Each step is at most $\mathcal{O}(L^3)$ in the angular indices. The input radial-channel count $N_{\rm in}$ (step 1a) and output radial-channel count $N_{\rm out}$ (step 4) are distinct knobs, both entering only through $L^2$ factors.}
\label{tab:structured-scaling}
\end{table}

The total cost of this structured pipeline scales as $\mathcal{O}(ICL^3 + ICL^2(2N_{\rm in}+N_{\rm out}) + ICL^2)$. The radial channel counts $N_{\rm in}$ and $N_{\rm out}$ enter only through $L^2$ factors, never through $L^3$, since applying both the radial CP factors $c^{(s)}_{cnl}$ (forward) and $\lambda_c^{(nl)}$ (backward) in SH space rather than on the grid decouples the dependence on radial channels from the angular transforms.

\subsection{Pipeline for the parity-aware message-passing coupling on a structured grid}
\label{sec:pipeline-msgpass}

This pipeline evaluates the parity-aware message-passing coupling Eq.\eqref{eq:msgpass-parity-start} on a structured Gauss--Legendre $\times$ uniform-Fourier grid, with parity-polar edge feature $E_{jin_1l_1m_1}=R_{n_1l_1}(r_{ji})\,Y_{l_1m_1}(\hat r_{ji})$ on each bond and parity-labeled node feature $I_{jn_2l_2m_2,p_2}$ on each atom. The factorized form of the edge feature, a radial function times a single spherical harmonic at the bond direction, is exploited by absorbing the radial CP factor into a single index over $(c, l_1)$ and avoiding materialization of a full SH-expanded tensor on the edge side. Two variants, the on-site bilinear contraction Eq.\eqref{eq:phi-cg} with general parity labels and the two-parity-bearing-fields coupling Eq.\eqref{eq:twoparity-phi}, lack this factorization on either side and require explicit forward transforms on both factors. Modifications are stated in the paragraph at the end of this subsection.

Following Sec.~\ref{sec:msgpass-parity-cases}, every output sector $(l, p)$ is the sum of two runs, namely a Category-$\alpha$ run with the direct Gaunt coupling at output degree $l$ on parity-resolved input slots, and a Category-$\beta$ run with the surface-curl coupling on the grid. Let $j$ denote the neighbor index and $c$ the CP rank. The angular prefactor is absorbed into an asymmetric CP decomposition of the learnable weights. For Category-$\alpha$ couplings ($l_1+l_2+l$ even), the prefactor is $f(l_1, l_2, l)$ of Eq.\eqref{eq:f-def},
\begin{equation}
\label{eq:msgpass-cp-alpha}
c_{n_1 l_1\, n_2 l_2}^{(nl,p,\alpha)}\,f(l_1, l_2, l) = \sum_c \lambda_c^{(nl,p,\alpha)}\,c^{(1,\alpha,p)}_{c n_1 l_1}\,c^{(2,\alpha,p)}_{c n_2 l_2}\,.
\end{equation}
For Category-$\beta$ couplings ($l_1+l_2+l$ odd), the prefactor is the parity-odd constant $\kappa(l_1, l_2, l)$ introduced through Eq.\eqref{eq:parity-odd-identity}, giving
\begin{equation}
\label{eq:msgpass-cp-beta}
c_{n_1 l_1\, n_2 l_2}^{(nl,p,\beta)}\,\kappa(l_1, l_2, l) = \sum_c \lambda_c^{(nl,p,\beta)}\,c^{(1,\beta,p)}_{c n_1 l_1}\,c^{(2,\beta,p)}_{c n_2 l_2}\,.
\end{equation}
Each line of Eqs.\eqref{eq:msgpass-cp-alpha}--\eqref{eq:msgpass-cp-beta} is a rank-$C$ CP decomposition of the prefactor-weighted coupling tensor that the model parameterizes directly. The two categories use independent CP weights and are summed at the end.

\paragraph{Category $\alpha$: direct Gaunt run.}
For each output parity sector $p \in \{+1,-1\}$, the slot rule
\begin{equation}
\label{eq:cat-alpha-slot}
p_2(l, l_2, p) \;=\; p \cdot (-1)^{l + l_2}
\end{equation}
depends on the output degree $l$ through $(-1)^l$, and a single source feature cannot simultaneously serve outputs of both $l$-parities. We split by output-$l$ parity $\sigma \in \{+1,-1\}$ and build two source features per output parity,
\begin{equation}
\label{eq:msgpass-Iappa}
I^{(\alpha,p,\sigma)}_{jn_2l_2m_2} \;=\; I_{jn_2l_2m_2,\,\sigma p (-1)^{l_2}}\,,
\end{equation}
running the pipeline below once for each $(p, \sigma)$, with the $\sigma = +1$ run feeding even-$l$ output sectors and the $\sigma = -1$ run feeding odd-$l$ sectors. This reproduces both Case~A (natural-parity input feeding natural output) and Case~C (parity-flipped input feeding parity-flipped output).

The resulting tensor $I^{(\alpha,p,\sigma)}_{jn_2l_2m_2}$ has no $p_2$ index and is fed into the structured-grid pipeline. The edge feature factors as $E = R(r_{ji})\,Y_{l_1m_1}(\hat r_{ji})$, so the radial contraction is applied to $R$ alone,
\begin{equation}
\label{eq:msgpass-tildeR}
\tilde R^{(\alpha,p,\sigma)}_{jicl_1} = \sum_{n_1} c^{(1,\alpha,p,\sigma)}_{c n_1 l_1}\,R_{n_1 l_1}(r_{ji})\,,
\end{equation}
\begin{equation}
\label{eq:msgpass-HI}
H^{(I,\alpha,p,\sigma)}_{j c l_2 m_2} = \sum_{n_2} c^{(2,\alpha,p,\sigma)}_{c n_2 l_2}\,I^{(\alpha,p,\sigma)}_{j n_2 l_2 m_2} \qquad (|m_2|\le l_2)\,.
\end{equation}
Avoiding materialization of an $H^{(E)}$ tensor with full $(l_1, m_1)$ dependence saves a factor of $L$ on the edge-radial work. The Legendre transform over $\theta$ then runs as
\begin{equation}
\label{eq:msgpass-FE}
F^{(E,\alpha,p,\sigma)}_{ji c m_1 u} = \sum_{l_1=|m_1|}^{l_{\max}} p_{u l_1 m_1}\,\tilde R^{(\alpha,p,\sigma)}_{jicl_1}\,Y_{l_1 m_1}(\hat r_{ji})\,,
\end{equation}
\begin{equation}
\label{eq:msgpass-FI}
F^{(I,\alpha,p,\sigma)}_{j c m_2 u} = \sum_{l_2=|m_2|}^{l_{\max}} p_{u l_2 m_2}\,H^{(I,\alpha,p,\sigma)}_{j c l_2 m_2}\,,
\end{equation}
followed by the Fourier transform over $\phi$,
\begin{equation}
\label{eq:msgpass-AE}
\mathcal{A}^{(E,\alpha,p,\sigma)}_{ji c uv} = \sum_{m_1=-l_{\max}}^{l_{\max}} q_{v m_1}\,F^{(E,\alpha,p,\sigma)}_{ji c m_1 u}\,,
\end{equation}
\begin{equation}
\label{eq:msgpass-AI}
\mathcal{A}^{(I,\alpha,p,\sigma)}_{j c uv} = \sum_{m_2=-l_{\max}}^{l_{\max}} q_{v m_2}\,F^{(I,\alpha,p,\sigma)}_{j c m_2 u}\,.
\end{equation}
The two grid fields are multiplied pointwise, with the neighbor summation pushed inside the spatial grid,
\begin{equation}
\label{eq:msgpass-S}
S^{(\alpha,p,\sigma)}_{i c uv} = \sum_j \mathcal{A}^{(E,\alpha,p,\sigma)}_{ji c uv}\,\mathcal{A}^{(I,\alpha,p,\sigma)}_{j c uv}\,.
\end{equation}
The pooled grid field is projected back to spherical-harmonic coefficients in CP-rank space by the inverse Fourier and Legendre transforms,
\begin{equation}
\label{eq:msgpass-Tmu}
T^{(\alpha,p,\sigma)}_{i c m u} = \frac{2\pi}{V} \sum_{v=1}^{V} q_{v m}^*\,S^{(\alpha,p,\sigma)}_{i c uv}\,,
\end{equation}
\begin{equation}
\label{eq:msgpass-Tlm}
T^{(\alpha,p,\sigma)}_{iclm} = \sum_{u=1}^{U} w_u\,p_{u l m}\,T^{(\alpha,p,\sigma)}_{i c m u} \qquad (|m|\le l,\ (-1)^l = \sigma)\,.
\end{equation}
Finally, the learnable channel projection reintroduces the output $(n,l)$ index,
\begin{equation}
\label{eq:cat-alpha-phi}
\varphi^{(\alpha)}_{inlm,p} \;=\; \sum_c \lambda^{(nl,p,\alpha)}_c\,T^{(\alpha,p,\sigma=(-1)^l)}_{iclm} \qquad (|m|\le l)\,.
\end{equation}

\paragraph{Category $\beta$: surface-curl run.}
For each output parity sector $p \in \{+1,-1\}$ and each output-$l$ parity $\sigma \in \{+1,-1\}$, build the source
\begin{equation}
\label{eq:cat-beta-slot}
I^{(\beta,p,\sigma)}_{jn_2l_2m_2} \;=\; I_{jn_2l_2m_2,\,-\sigma p (-1)^{l_2}}\,,
\end{equation}
the opposite slot of the Category-$\alpha$ rule Eq.\eqref{eq:msgpass-Iappa}. This reproduces both Case~B and Case~D.

Run the forward transforms Eq.\eqref{eq:msgpass-tildeR}--Eq.\eqref{eq:msgpass-AI} on $I^{(\beta,p,\sigma)}$ and on the edge field, with CP factors $c^{(1,\beta,p,\sigma)}_{cn_1l_1}, c^{(2,\beta,p,\sigma)}_{cn_2l_2}$ fitted independently. This yields the grid-evaluated tensors $\mathcal{A}^{(E,\beta,p,\sigma)}_{ji c uv}$ and $\mathcal{A}^{(I,\beta,p,\sigma)}_{j c uv}$.

Compute the surface-curl Jacobian on the grid using Eq.\eqref{eq:poisson-S2-x} in the $x = \cos\theta$ coordinate, which avoids $1/\sin\theta$, and
\begin{multline}
\label{eq:cat-beta-curl}
S^{(\beta,p,\sigma)}_{i c uv} \;=\; \sum_j \Bigl(\partial_\phi\mathcal{A}^{(E,\beta,p,\sigma)}_{ji c uv}\,\partial_x\mathcal{A}^{(I,\beta,p,\sigma)}_{j c uv} \\
\;-\; \partial_x\mathcal{A}^{(E,\beta,p,\sigma)}_{ji c uv}\,\partial_\phi\mathcal{A}^{(I,\beta,p,\sigma)}_{j c uv}\Bigr)\,.
\end{multline}
Since $\partial_\phi Y_{lm} = im\,Y_{lm}$, the $\partial_\phi$ derivative is free, obtained from the standard Legendre output by one extra Fourier pass with $im\,q_{vm}$. The $\partial_x$ derivative requires one additional Legendre pass with the precomputed transform $\partial_x p_{ulm}$ at the same $\mathcal{O}(L^3)$ cost as the ordinary Legendre transform. The integrand flips sign under $1\leftrightarrow 2$ exchange by the antisymmetry of the $2\times 2$ determinant, matching the parity-odd CG. We then run the inverse transforms Eq.\eqref{eq:msgpass-Tmu}--Eq.\eqref{eq:msgpass-Tlm} on this scalar field, restricting the output to $(-1)^l = \sigma$, and apply the channel projection,
\begin{equation}
\label{eq:cat-beta-phi}
\varphi^{(\beta)}_{inlm,p} \;=\; \sum_c \lambda^{(nl,p,\beta)}_c\,T^{(\beta,p,\sigma=(-1)^l)}_{iclm} \qquad (|m|\le l)\,.
\end{equation}

\paragraph{Final assembly.}
The full target feature is the sum of the Category-$\alpha$ and Category-$\beta$ contributions.
\begin{equation}
\label{eq:msgpass-total}
\varphi_{inlm,p} \;=\; \varphi^{(\alpha)}_{inlm,p} \;+\; \varphi^{(\beta)}_{inlm,p} \qquad (|m|\le l)\,.
\end{equation}

\paragraph{Computational scaling.}
Let $I_{\text{nodes}}$ be the number of atoms, $Z$ the average number of neighbors per atom, $C$ the CP rank, $L = l_{\max}+1$, and $N_{\rm out}$ the output radial-channel count. Each Category-$\alpha$ run costs $\mathcal{O}(I_{\text{nodes}}[ZCL^3 + ZCL^2 + N_{\rm out}CL^2])$ for the dominant terms (edge transform, per-bond grid-pointwise product, and output channel projection). Each Category-$\beta$ run requires one additional Legendre pass with $\partial_x p_{ulm}$ for the $\partial_x$ component, giving $\mathcal{O}(I_{\text{nodes}}[2ZCL^3 + ZCL^2 + N_{\rm out}CL^2])$. With one $\alpha$ run and one $\beta$ run per $(p, \sigma)$ combination, two output parities, and two output-$l$ parities per output parity,
\begin{equation}
\label{eq:msgpass-total-cost}
\mathrm{cost}_{\rm total} \;=\; 12\,\mathcal{O}\bigl(I_{\text{nodes}}\,[\,Z C L^3 + Z C L^2 + N_{\rm out} C L^2\,]\bigr)\,,
\end{equation}
i.e.\ a constant factor of approximately $12\times$ the parity-unaware pipeline cost for full coverage of all CG channels in both parity sectors of distinct-factor couplings. The asymptotic algorithmic cost is $\mathcal{O}(L^3)$, but the $ZCL^2$ per-bond pointwise term carries the same $ZC$ prefactor as the leading edge-transform term, and the $L^2$-to-$L^3$ crossover is at $L \sim 1$ and the empirical wall-clock scales as $L^2$ across the practical $l_{\max}$ range. See Sec.~\ref{sec:empirical-scaling} for the benchmark interpretation.

The $12\times$ count is a conservative upper bound that treats each $(p, \sigma)$ run as fully independent. The four source constructions of Eq.\eqref{eq:msgpass-Iappa} use only two distinct $l_2$-parity selections of the slots $I_{\pm}$, so an implementation that shares the forward transforms of $I_{\pm}$ and $E$ across $(p, \sigma)$ combinations and amortizes the $\partial_x$ Legendre passes between Categories $\alpha$ and $\beta$ is expected to bring the effective constant factor closer to $\sim 8\times$.

For self-coupled bilinear contractions ($A^{(1)} = A^{(2)} = A$) with symmetric channel-mixing, Category $\beta$ vanishes by Eq.\eqref{eq:self-coupled-vanish} and the cost reduces to two Category-$\alpha$ runs per output parity. (For self-coupled inputs with antisymmetric channel mixing, as may arise in magnetic or tensorial properties~\cite{drautz2020}, Category $\beta$ contributes non-trivially, as verified in Appendix~\ref{sec:numerical-validation}.)

\paragraph{On-site bilinear contraction with general parity.}
For the on-site contraction Eq.\eqref{eq:phi-cg} with general parity-labeled inputs, the pipeline above carries over with three changes. First, both factors are on-site, so the neighbor index $j$ and summation $\sum_j$ are dropped, and the grid product Eq.\eqref{eq:msgpass-S} becomes $S^{(\alpha,p,\sigma)}_{icuv} = \mathcal{A}^{(1,\alpha,p,\sigma)}_{icuv}\,\mathcal{A}^{(2,\alpha,p,\sigma)}_{icuv}$, and likewise for Category $\beta$. Second, neither factor has the $R\cdot Y$ factorization, so both run through the structured-grid forward pipeline of Sec.~\ref{sec:pipeline-structured}. Third, parity dispatch operates on both factors. For each output sector $p$, two pairs $(p_1, p_2) \in \{(+1, p),\,(-1,-p)\}$ satisfy $p_1 p_2 = p$. The cost per run is $\mathcal{O}(I[CL^3 + NCL^2])$ (no $Z$ factor). With two pairs $\times$ two output sectors $\times$ two output-$l$ parities $\times$ ($1\alpha + 2\beta$) runs per combination, the upper-bound cost is $24\,\mathcal{O}(I[CL^3 + NCL^2])$, with the same forward-transform sharing as above expected to reduce the effective factor to roughly half. The asymptotic $\mathcal{O}(L^3)$ scaling is preserved.

\paragraph{Two parity-bearing fields.}
For the coupling Eq.\eqref{eq:twoparity-phi} of Sec.~\ref{sec:two-parity}, the pipeline carries over with two changes. The dispatch reads from both $I^{(E)}$ and $I^{(N)}$ with independent parity-slot selections at $(l_1, p_1)$ and $(l_2, p_2)$. Since $I^{(E)}$ no longer factors as $R(r_{ji})\,Y(\hat r_{ji})$, an explicit radial contraction is required on the edge side, replacing $\tilde R$ by an analogue of Eq.\eqref{eq:msgpass-HI}. The surface-curl Eq.\eqref{eq:cat-beta-curl} acts on the two grid fields in direct analogy to the single-parity-input case. With four input-parity combinations, two output-$l$ parities, and Categories~$\alpha$ and~$\beta$, the upper-bound cost is approximately
\begin{equation}
\label{eq:twoparity-total-cost}
24\,\mathcal{O}\bigl(I_{\text{nodes}}\,[\,Z C L^3 + Z N C L^2 + N C L^2\,]\bigr)\,,
\end{equation}
again reducible by forward-transform sharing as discussed for the message-passing case.

\begin{figure*}[tbp]
\centering
\includegraphics[width=\textwidth]{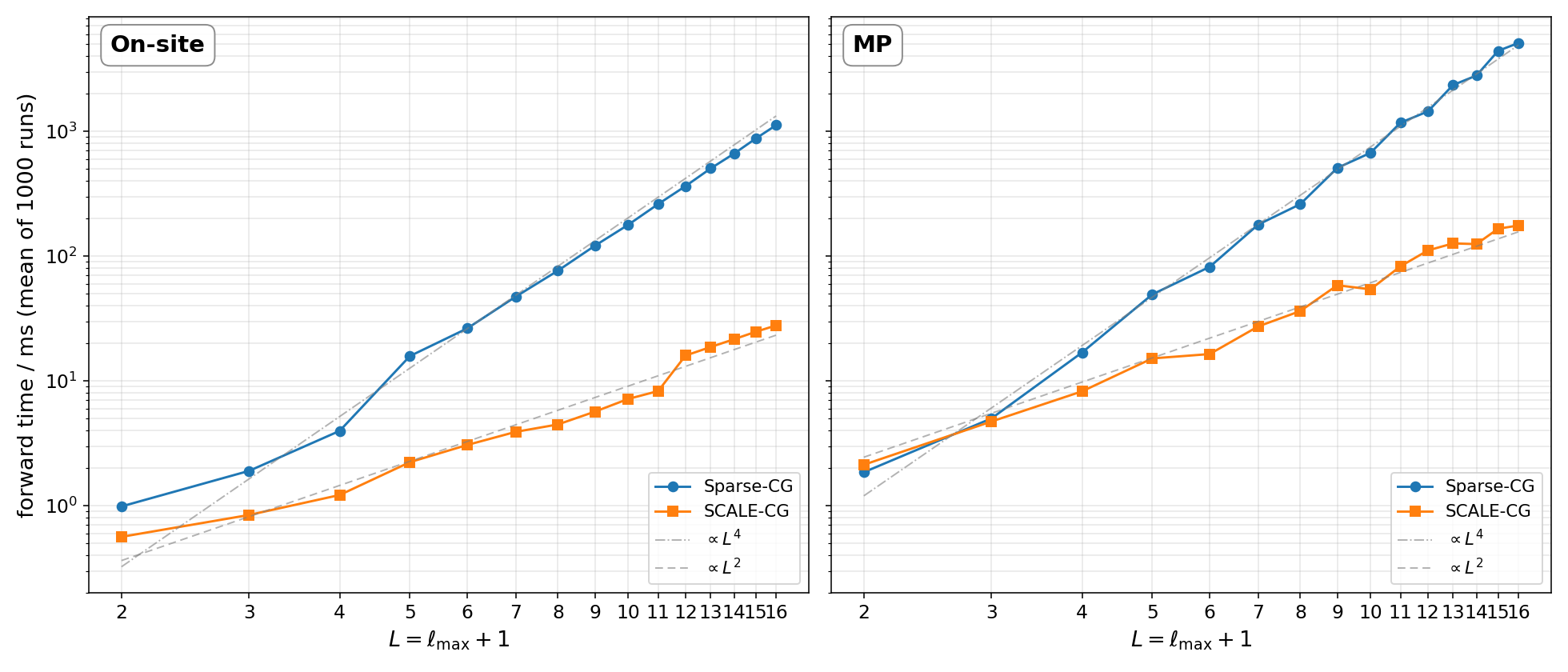}
\caption{Forward-pass wall time vs. $L = \ell_{\max}+1$ for the on-site contraction Eq.~\eqref{eq:phi-cg} (left)
  and the message-passing coupling Eq.~\eqref{eq:msgpass-parity-start} (right). Blue: optimized sparse 
 direct CG sum. Orange: SCALE-CG. Dashed grey: best-fit reference lines $\propto L^4$ (sparse) and       
 $\propto L^2$ (SCALE-CG). }
\label{fig:scaling}
\end{figure*}

\section{Numerical performance}
\label{sec:empirical-scaling}

We benchmark SCALE-CG for the atomic-base contraction Eq.~\eqref{eq:phi-cg}, restricted to natural parity (left panel of Fig.~\ref{fig:scaling}), and message passing Eq.~\eqref{eq:msgpass-parity-start} with full parity (right panel), at CP rank $C = 64$, $N = 64$ radial channels, with input and output bandlimits set equal at $L = \ell_{\max}+1$. Each point is evaluated for $N_{\mathrm{atoms}} = 1000$ random atomic environments, with $50$ neighbors per atom in the message-passing setting, and each forward pass is timed over $1000$ calls. The curves show the mean. The straight reference lines $\propto L^{4}$ and $\propto L^{2}$ are best-fit log-linear intercepts (constraining the slope to the indicated exponent) to the corresponding measured curve. SCALE-CG is implemented in the \texttt{GRACEmaker}~\cite{bochkarev2024, gracemaker} package using TensorFlow~\cite{tensorflow2015-whitepaper} $2.20$ with XLA, and all measurements were obtained on a single NVIDIA H200 GPU with FP32 precision. In the present implementation the azimuthal forward and inverse transforms are evaluated as dense matrix multiplications rather than fast Fourier transforms, giving an effective $L^3$ rather than $L^2\log L$ for those stages without affecting the leading asymptote.

Empirically, SCALE-CG scales as $L^2$ rather than its asymptotic $L^3$ FLOP-count prediction. The optimized sparse-CG sum also scales at $L^4$ rather than $L^5$. On the GPU hardware used here, the wall-clock is set by memory traffic on grid tensors of size $\propto L^2$, not by the underlying $L^3$ arithmetic. For SCALE-CG, the heaviest intermediate is the per-bond grid field of shape $[B, C, U, V]$ (per-bond batch $B = IZ$ for message passing, per-atom batch $B = I$ for on-site), with memory that scales as $L^2$ from the grid dimensions $UV \approx \tfrac{9}{2}L^2$. The Legendre and Fourier transforms carry $L^3$ arithmetic but must read and write the same $L^2$-sized tensor and are therefore memory-bandwidth-bound. A per-step breakdown of the message-passing pipeline confirming this picture is given in Appendix~\ref{sec:profiling-appendix}.

Based on a pure operation count, a direct Clebsch--Gordan implementation that exploits full sparsity through $m_1 + m_2 = m$ and the triangle inequality has a leading multiplication count $\tfrac{4}{15}L^5$ for the parity-complete operation, halved within a single parity sector. The structured-grid algorithm of Sec.~\ref{sec:pipeline-structured} on a grid of $U \approx \tfrac{3}{2}L$ in $\theta$ and $V \approx 3L$ in $\phi$ gives approximately $10L^3$ multiplications across the dominant Legendre and Fourier stages of one forward and inverse pass. Equating these leading counts places the direct-vs-SCALE-CG crossover at $l_{\max} \approx 8$. However, both the on-site and the message-passing operations contain sub-leading $L^2$ contributions in addition to the leading $L^3$. For the on-site contraction, $\mathcal{O}(IC(2N_{\rm in}+N_{\rm out})L^2)$ radial-channel projections and an $\mathcal{O}(ICL^2)$ grid-pointwise term accompany the $\mathcal{O}(ICL^3)$ Legendre transforms (Table~\ref{tab:structured-scaling}). For message passing, the $\mathcal{O}(ZCL^2)$ per-bond grid-pointwise term shares the $ZC$ prefactor of the leading $\mathcal{O}(ZCL^3)$ edge-transform. For the on-site contraction these sub-leading contributions dominate up to the algorithmic crossover at $L \approx 2N_{\rm in}+N_{\rm out}$, well above the range used in atomistic models. For message passing the shared $ZC$ prefactor reduces the algorithmic crossover to $L \approx 1$, and the empirical $L^2$ persistence is set instead by the bandwidth ceiling discussed above.

\section{Summary and Conclusions}
\label{sec:summary}

The full uncontracted CGTP, generating independent $(l_1, l_2, l)$ outputs, is bounded below by $\mathcal{O}(L^4)$~\cite{xie2026}. We present SCALE-CG, an efficient algorithm that avoids this bottleneck for the contracted CP-factorized Clebsch--Gordan tensor products of ACE-family equivariant potentials, taking it down to $\mathcal{O}(L^3)$. By mapping the Clebsch--Gordan integral to a discrete summation over a structured Gauss--Legendre and Fourier tensor-product grid, we separated the polar and azimuthal variables. This factors out the radial channel projections and bounds the cost of any single forward or inverse transform step by $\mathcal{O}(L^3)$.

We resolved the algebraic limitation of scalar grids, namely their inability to compute the antisymmetric parity-odd Clebsch--Gordan channels via commutative scalar multiplication, by replacing the symmetric pointwise product with the antisymmetric surface-curl pairing $\hat{r} \cdot [\nabla_{\mathrm{S}^2} A \times \nabla_{\mathrm{S}^2} B]$, equivalently the spherical Poisson bracket $\{A, B\}_{S^2}$.

Initial benchmarks show $L^2$ scaling and improved efficiency of SCALE-CG over an optimized sparse direct CG sum for all $L$. SCALE-CG is directly applicable to existing CP-factorized $\mathrm{E}(3)$-equivariant potentials of the ACE-family, and the reduced asymptotic scaling enables exploration of larger angular momentum cutoffs at fixed channel rank.

\subsection*{AUTHOR CONTRIBUTIONS}

R.D. conceived the study, developed and implemented the algorithm, and wrote the original draft. A.B. and Y.L. carried out the timing and scaling benchmarks. All authors contributed to reviewing and editing the manuscript.

\begin{acknowledgments}
Derivations and exposition were prepared with substantial assistance from the large language models Claude (Anthropic), Gemini (Google), and ChatGPT (OpenAI), used as adversarial reviewers and formal-derivation aids. A.B. and R.D. acknowledge financial support by Deutsche Forschungsgemeinschaft (DFG) through project 409476157 (A06).
\end{acknowledgments}

\appendix

\section{Proof of the parity-odd identity}
\label{sec:appendix-derivation}

We prove that the surface-curl integral
\begin{multline}
\label{eq:Q-def-appendix}
Q^{lm}_{l_1 m_1\, l_2 m_2} \;\equiv\; \int d\Omega\, Y_{lm}^*(\hat r) \\
\times \hat r \cdot \bigl[\nabla_{S^2} Y_{l_1 m_1}(\hat r) \times \nabla_{S^2} Y_{l_2 m_2}(\hat r)\bigr]
\end{multline}
satisfies $Q^{lm}_{l_1 m_1\, l_2 m_2} = \kappa^{-1}(l_1, l_2, l)\,G^{lm}_{l_1 m_1\, l_2 m_2}$ for a single $(m_1, m_2, m)$-independent scalar $\kappa^{-1}(l_1, l_2, l)$, with $\kappa^{-1} = 0$ on the parity-even support $l_1 + l_2 + l$ even and $\kappa^{-1} \neq 0$ on the parity-odd support $l_1 + l_2 + l$ odd. The proof has four steps: (i)~establish $\mathrm{SO}(3)$ covariance of $Q$, (ii)~apply Schur's lemma in the form of the Wigner--Eckart theorem~\cite{varshalovich1988} to constrain $Q$ to be proportional to the Clebsch--Gordan coefficient, (iii)~use inversion parity to restrict the support to $l_1 + l_2 + l$ odd, and (iv)~derive a closed-form expression for $\kappa^{-1}$ on the parity-odd support.

\subsection*{Step 1: $\mathrm{SO}(3)$ covariance}

Let $R \in \mathrm{SO}(3)$ act on functions on $S^2$ by $(R\!\cdot\! f)(\hat r) = f(R^{-1}\hat r)$. The surface gradient is rotation-equivariant, $\nabla_{S^2}(R\!\cdot\! f)(\hat r) = R\,[\nabla_{S^2} f](R^{-1}\hat r)$ (chain rule), and rotations preserve cross products and respect the radial projection, $(Ru)\times(Rv) = R(u\times v)$ and $\hat r\cdot(Rw) = (R^{-1}\hat r)\cdot w$ on $S^2$. Combining these, the bilinear form $f[A, B](\hat r) = \hat r \cdot [\nabla_{S^2} A \times \nabla_{S^2} B]$ behaves as a (pseudo-)scalar on $S^2$ in its functional arguments, $f[R\!\cdot\! A, R\!\cdot\! B](\hat r) = f[A, B](R^{-1}\hat r)$.

The spherical harmonics carry the $\mathrm{SO}(3)$ representation $D^l$,
\begin{equation}
\label{eq:Y-rotation}
Y_{lm}(R^{-1}\hat r) = \sum_{m'} D^l_{m m'}(R^{-1})\,Y_{lm'}(\hat r)\,.
\end{equation}
Substituting Eq.\eqref{eq:Y-rotation} into Eq.\eqref{eq:Q-def-appendix} and changing variables $\hat r \to R\hat r$ (with $d\Omega$ invariant) gives the covariance identity
\begin{multline}
\label{eq:Q-cov}
\sum_{m_1' m_2' m'} D^{l_1}_{m_1 m_1'}(R)\, D^{l_2}_{m_2 m_2'}(R)\, [D^l_{m m'}(R)]^* \\
\times Q^{lm'}_{l_1 m_1'\, l_2 m_2'} \;=\; Q^{lm}_{l_1 m_1\, l_2 m_2}
\end{multline}
for every $R \in \mathrm{SO}(3)$.

\subsection*{Step 2: Schur's lemma fixes $Q \propto G$}

Eq.\eqref{eq:Q-cov} states that $Q$ is an $\mathrm{SO}(3)$-equivariant linear map $T_Q: V_{l_1} \otimes V_{l_2} \to V_l$. By the Clebsch--Gordan decomposition,
\begin{equation}
\label{eq:CG-decomp}
V_{l_1} \otimes V_{l_2} \;\cong\; \bigoplus_{l' = |l_1 - l_2|}^{l_1 + l_2} V_{l'}\,,
\end{equation}
$V_l$ appears with multiplicity one inside the triangle range and zero outside, so by Schur's lemma the space of equivariant maps $V_{l_1} \otimes V_{l_2} \to V_l$ is one-dimensional in the triangle range. The CG coefficient $G^{lm}_{l_1 m_1\, l_2 m_2}$ is the matrix element of the canonical equivariant projector onto $V_l$ and spans this one-dimensional space, so by the Wigner--Eckart theorem
\begin{equation}
\label{eq:Q-prop-G}
Q^{lm}_{l_1 m_1\, l_2 m_2} \;=\; \kappa^{-1}(l_1, l_2, l)\,G^{lm}_{l_1 m_1\, l_2 m_2}\,,
\end{equation}
with $\kappa^{-1}$ depending only on $(l_1, l_2, l)$.

\subsection*{Step 3: Inversion parity fixes the support}

Schur's lemma forces $Q$ to be proportional to $G$, but does not yet identify on which $(l_1, l_2, l)$ triples the proportionality constant $\kappa^{-1}$ vanishes. The cleanest determination comes from inversion parity. Under spatial inversion $\hat r \to -\hat r$, the spherical harmonic $Y_{lm}$ picks up a factor $(-1)^l$. The surface gradient $\nabla_{S^2}$ acts on a tangent vector field, so $\nabla_{S^2} Y_{lm}(-\hat r) = -(-1)^{l}\,\nabla_{S^2} Y_{lm}(\hat r)$ (the extra $-1$ comes from the orientation flip of the tangent plane at $-\hat r$ relative to $\hat r$). Combining the two factors and the radial vector $\hat r$ which itself flips sign, the integrand of Eq.\eqref{eq:Q-def-appendix} transforms as
\begin{multline}
\label{eq:integrand-parity}
\hat r \cdot [\nabla_{S^2} Y_{l_1m_1} \times \nabla_{S^2} Y_{l_2m_2}]\,Y^*_{lm} \;\xrightarrow{\hat r \to -\hat r}\; \\
(-1)^{l_1+l_2+l+1}\,\hat r \cdot [\nabla_{S^2} Y_{l_1m_1} \times \nabla_{S^2} Y_{l_2m_2}]\,Y^*_{lm}\,.
\end{multline}
The integral over the full sphere of any function with parity $-1$ under $\hat r \to -\hat r$ vanishes, so
\begin{equation}
\label{eq:Q-parity-vanish}
Q^{lm}_{l_1 m_1\, l_2 m_2} \;=\; 0 \qquad \text{whenever}\ l_1 + l_2 + l\ \text{is even.}
\end{equation}
$\kappa^{-1}(l_1, l_2, l) = 0$ on the parity-even support, leaving only the parity-odd support $l_1 + l_2 + l\ \text{odd}$.

\subsection*{Step 4: Closed-form expression on the parity-odd triangle}

We give the explicit closed form of $\kappa^{-1}$ by expanding the surface gradient in the basis of spin-weighted spherical harmonics ${}_s Y_{lm}$~\cite{newmanPenrose1966,goldberg1967}. With $\mathbf{m} = (\hat{\mathbf{e}}_\theta + i\hat{\mathbf{e}}_\phi)/\sqrt{2}$ and the Goldberg sign convention~\cite[\S10]{varshalovich1988},
\begin{equation}
\label{eq:gradient-VSH}
\nabla_{S^2} Y_{lm} \;=\; \sqrt{\frac{l(l+1)}{2}}\bigl(\,{}_1 Y_{lm}\,\mathbf{m}^* \;-\; {}_{-1} Y_{lm}\,\mathbf{m}\,\bigr)\,.
\end{equation}
The cross product of the basis vectors gives $\mathbf{m}^* \times \mathbf{m} = i\hat r$, so the radial-projected surface curl reduces to a difference of products of spin-weighted harmonics.
\begin{multline}
\label{eq:gradient-cross-VSH}
\hat r \cdot [\nabla_{S^2} Y_{l_1 m_1} \times \nabla_{S^2} Y_{l_2 m_2}] \\
=\; -i\,\frac{\sqrt{l_1(l_1+1)\,l_2(l_2+1)}}{2} \\
\times\,\bigl[\,{}_1 Y_{l_1 m_1}\,{}_{-1} Y_{l_2 m_2} \;-\; {}_{-1} Y_{l_1 m_1}\,{}_1 Y_{l_2 m_2}\,\bigr]\,.
\end{multline}
Multiplying by $Y^*_{lm} = (-1)^m\,{}_0 Y_{l, -m}$ and integrating over the sphere invokes the standard identity for the integral of three spin-weighted harmonics, evaluating each term to a product of two Wigner 3-$j$ symbols. The bottom rows of the spin-3-$j$ for the two terms are $(0, -1, 1)$ and $(0, 1, -1)$, related by sign reversal. Sign reversal of the bottom row of a 3-$j$ symbol multiplies it by $(-1)^{l_1+l_2+l}$, which equals $-1$ on the parity-odd support, so the two terms add constructively. Factoring out the Clebsch--Gordan coefficient via the standard 3-$j$/CG relation gives
\begin{multline}
\label{eq:kappa-inv-closed}
\kappa^{-1}(l_1, l_2, l) \;=\; -i\,(-1)^{l_1-l_2} \\
\times\;\sqrt{\frac{(2l_1+1)(2l_2+1)\,l_1(l_1+1)\,l_2(l_2+1)}{4\pi}} \\
\times\,\begin{pmatrix} l_1 & l_2 & l \\ -1 & 1 & 0 \end{pmatrix}\,.
\end{multline}
The Wigner 3-$j$ symbol $\bigl(\begin{smallmatrix} l_1 & l_2 & l \\ -1 & 1 & 0 \end{smallmatrix}\bigr)$ obeys the standard triangle inequality $|l_1-l_2| \le l \le l_1+l_2$ and is non-vanishing whenever $l_1+l_2+l$ is odd. Together with the prefactor $\sqrt{l_1(l_1+1)\,l_2(l_2+1)}$ which forces $l_1, l_2 \ge 1$, this gives $\kappa^{-1}(l_1, l_2, l) \neq 0$ on the full parity-odd support $l_1, l_2 \ge 1$, $|l_1-l_2| \le l \le l_1+l_2$, $l_1+l_2+l$ odd. Closed-form integral expressions for the antisymmetric Gaunt analogue underlying $\kappa^{-1}$ have also been obtained by Heyraud et al.~\cite{heyraud2026} via direct simplification of the vector-spherical-tensor-product of Xie et al.~\cite{xie2026}, giving a result equivalent to our Eq.~\eqref{eq:kappa-inv-closed} on the parity-odd support.

For the canonical example $(l_1, l_2, l) = (1, 1, 1)$, take $m_1 = +1$, $m_2 = -1$, $m = 0$. Using the standard convention $Y_{1,\pm 1} = \mp\sqrt{3/(8\pi)}\sin\theta\,e^{\pm i\phi}$ and $Y_{1,0} = \sqrt{3/(4\pi)}\cos\theta$, the surface-curl Jacobian Eq.\eqref{eq:poisson-S2} evaluates to
\begin{equation}
\frac{1}{\sin\theta}(\partial_\theta Y_{1,1}\,\partial_\phi Y_{1,-1} - \partial_\phi Y_{1,1}\,\partial_\theta Y_{1,-1}) \;=\; \frac{3i}{4\pi}\cos\theta\,,
\end{equation}
so that
\begin{equation}
Q^{1,0}_{1,1\,1,-1} \;=\; \int Y^*_{1,0}(\hat r)\,\frac{3i}{4\pi}\cos\theta\,d\Omega \;=\; i\sqrt{\frac{3}{4\pi}}\,.
\end{equation}
The corresponding Clebsch--Gordan coefficient is $G^{1,0}_{1,1\,1,-1} = +1/\sqrt{2}$, giving
\begin{equation}
\label{eq:kappa-inv-111}
\kappa^{-1}(1, 1, 1) \;=\; \frac{Q^{1,0}_{1,1\,1,-1}}{G^{1,0}_{1,1\,1,-1}} \;=\; i\sqrt{\frac{3}{2\pi}} \;\ne\; 0\,,
\end{equation}
in agreement with Eq.\eqref{eq:kappa-inv-closed} evaluated at $(l_1, l_2, l) = (1, 1, 1)$ and matching the numerical extraction of Appendix~\ref{sec:numerical-validation} ($+0.690988\,i$) to eleven decimal places. Inverting Eq.\eqref{eq:Q-prop-G} on the parity-odd support gives the parity-odd identity Eq.\eqref{eq:parity-odd-identity}.

The proportionality constants $\kappa^{-1}(l_1, l_2, l)$ are the structure constants of the spherical Poisson algebra of spherical harmonics. Writing the bracket Eq.\eqref{eq:poisson-S2} as $\{Y_{l_1m_1}, Y_{l_2m_2}\}_{S^2}$ and expanding in SH on both sides,
\begin{equation}
\label{eq:poisson-structure-constants}
\{Y_{l_1m_1}, Y_{l_2m_2}\}_{S^2} \;=\; \sum_{l, m}\,\kappa^{-1}(l_1, l_2, l)\,G^{lm}_{l_1m_1\,l_2m_2}\,Y_{lm}\,,
\end{equation}
so the spherical harmonics close under $\{\cdot, \cdot\}_{S^2}$ into a Lie algebra with structure constants $\kappa^{-1}\,G$ on the parity-odd triangle.

\section{Numerical validation}
\label{sec:numerical-validation}

The surface-curl identity Eq.\eqref{eq:parity-odd-identity} and the parity-aware pipeline of Sec.~\ref{sec:pipeline-msgpass} are validated by direct numerical quadrature against the analytic Clebsch--Gordan coefficients computed by sympy~\cite{sympy}. All tests use complex spherical harmonics in the standard Condon--Shortley convention, integrated on a Gauss--Legendre $\times$ uniform-Fourier grid sized to be exact for polynomials of total degree $\ge 3\,\max(l_1, l_2, l)$, matching the bandwidth of the surface-curl integrand $Y_{lm}^*\,\hat r\cdot[\nabla Y_{l_1 m_1} \times \nabla Y_{l_2 m_2}]$ ($\le l_1 + l_2 + l$). Surface gradients use the analytic $\partial_x \tilde P_l^{|m|}(x)$ and $\partial_\phi Y_{lm} = im\,Y_{lm}$, the same expressions as in the runtime algorithm. Quoted error bars are dominated by the quadrature.

\subsection{Parity-odd identity recovery}

For each parity-odd triple $(l_1, l_2, l)$ we evaluate Eq.\eqref{eq:parity-odd-identity} on every $(m_1, m_2, m)$ entry with $m = m_1 + m_2$ and extract $\kappa^{-1}(l_1, l_2, l) = \mathrm{integral}/G^{lm}_{l_1m_1\,l_2m_2}$ from the non-vanishing entries. Three properties are checked simultaneously: (i) $\kappa^{-1}$ is purely imaginary in the complex basis, (ii) $\kappa^{-1}$ is independent of $(m_1, m_2, m)$, and (iii) the spread of $\kappa^{-1}$ across the $(m_1, m_2, m)$ entries is at the quadrature precision. We tested all parity-odd triples from $(1,1,1)$ to $(4,4,7)$. In every case $\kappa^{-1}$ was purely imaginary with spread $\le 10^{-10}$ across the $6$--$72$ non-vanishing entries per triple. The right-hand side reproduces $G^{lm}_{l_1m_1\,l_2m_2}$ to the same precision. The $(1, 1, 1)$ case is the canonical cross-product, with every non-vanishing entry yielding $\kappa^{-1} = +0.690988\,i$ to eleven decimal places, matching the analytic value Eq.\eqref{eq:kappa-inv-111}.

\subsection{Parity-even vanishing}

The same quadrature on parity-even triples ($l_1 + l_2 + l$ even) returns zero to numerical precision, since the parity-even CG channels are symmetric under $1\leftrightarrow 2$ while the surface-curl integrand is antisymmetric and the two cannot mix. Across all parity-even triples up to $(4,4,8)$, the maximum integral was $\sim 10^{-10}$, set by the quadrature.

\subsection{Antisymmetry under $1\leftrightarrow 2$ exchange}

The surface-curl integrand $\hat r \cdot [\nabla_{S^2} Y_{l_1m_1} \times \nabla_{S^2} Y_{l_2m_2}]$ flips sign exactly under exchange of the two factors. Direct evaluation across all parity-odd triples up to $(4,4,7)$ confirms this to machine precision, with the absolute value of the sum $\mathrm{integral}_{1,2} + \mathrm{integral}_{2,1}$ consistently below $10^{-17}$, i.e.\ at the level of double-precision roundoff rather than the quadrature precision.

\subsection{End-to-end consistency on random synthetic fields}

To verify that the surface-curl construction reproduces the contracted CG output not only on plain spherical harmonics but on arbitrary input fields, we generate independent random complex-Gaussian SH coefficients $A^{(1)}_{l_1 m_1}$, $A^{(2)}_{l_2 m_2}$, build the corresponding fields $A^{(1)}(\hat r), A^{(2)}(\hat r)$ on the grid, and compare two independent computations of the bilinear contraction $\varphi_{lm} = \sum_{m_1m_2} G^{lm}_{l_1m_1\,l_2m_2}\,A^{(1)}_{l_1m_1}\,A^{(2)}_{l_2m_2}$: (i) the direct CG sum, evaluated with the sympy~\cite{sympy} CG coefficients; and (ii) the surface-curl recovery via Eq.\eqref{eq:parity-odd-identity}, with $\kappa^{-1}(l_1, l_2, l)$ extracted once per $(l_1, l_2, l)$ from a single reference $(m_1, m_2, m)$ entry. Across $5$ independent random trials per $(l_1, l_2, l)$ triple and all output $m$, the two evaluations agree at the level of the quadrature precision, with $\max_m|\varphi_{lm}^{\text{direct}} - \varphi_{lm}^{\text{curl}}| \le 5.3 \times 10^{-10}$ across all parity-odd triples up to $(3,3,5)$.

\subsection{$\mathrm{SO}(3)$ equivariance under random rotations}

To confirm equivariance, we draw a uniformly random rotation $R \in \mathrm{SO}(3)$, transform the input coefficients by $A^{(i)}_{l_im_i} \to \sum_{m_i'} D^{l_i}_{m_im_i'}(R)\,A^{(i)}_{l_im_i'}$, and verify that the output transforms as the corresponding $D^l(R)$. Across all parity-odd triples up to $(3,3,5)$, $\max_m|\varphi^{\text{rot}}_{lm} - [D^l(R)\,\varphi^{\text{orig}}]_{lm}| \le 6.6 \times 10^{-10}$, at quadrature precision. Composition through quadrature on a fixed grid could in principle introduce small equivariance violations. The data show this does not occur.

\subsection{Self-coupled vanishing under symmetric channel-mixing}

To validate Eq.\eqref{eq:self-coupled-vanish}, we test the contracted parity-odd output with two channel matrices.
\begin{multline*}
\Phi^{(\text{sym/anti})}_{lm} = \sum_{n_1, n_2} M^{(\text{sym/anti})}_{n_1 n_2} \\
\times \sum_{m_1m_2} G^{lm}_{l_1m_1\,l_2m_2}\,A_{n_1l_1m_1}\,A_{n_2l_2m_2}
\end{multline*}
with $A_{nlm}$ random Gaussian, $l_1 = l_2$, $M^{\text{sym}}$ symmetric, $M^{\text{anti}}$ antisymmetric. Across all triples up to $(3,3,5)$, the symmetric-mixing channel vanishes at double-precision roundoff ($\sim 10^{-16}$) while the antisymmetric channel is order unity. The vanishing is exact and is a property of the channel-mixing symmetry rather than of the surface-curl construction.

\subsection{Parity-indexed message-passing coupling [Eq.\eqref{eq:msgpass-parity-start}]}
\label{sec:numerical-msgpass-parity}

We apply the surface-curl to the parity-indexed coupling Eq.\eqref{eq:msgpass-parity-start}, comparing (i)~the direct CG sum with explicit slot rule $p_2 = p\,(-1)^{l_1}$ and Category-$\alpha$/$\beta$ dispatch, and (ii)~the structured-grid pipeline of Sec.~\ref{sec:pipeline-msgpass}. The setup uses one bond at fixed $(\theta_e, \phi_e)$, one radial channel, $l_{\max} = 2$, random $R_{nl}(r_{ji})$, and random parity-labeled $I_{j n_2 l_2 m_2, p_2}$ in both parity slots.

To check the dispatch rule $p = (-1)^{l_1} p_2$, we restrict $R$ to a single edge degree and populate only one parity slot of $I$. With $l_1=0$ and $I$ on $p_2 = +1$, every $p_{\rm out}=+1$ output sector receives a non-zero contribution while $p_{\rm out}=-1$ vanishes exactly. Switching to $l_1=1$ flips this. Vanishing is exact at every output index.

With both parity slots populated and the full sum active, the surface-curl pipeline reproduces the direct CG sum across every output sector, with $l_{\rm out}=0$ at machine precision, $l_{\rm out}\ge 1$ at the quadrature precision $\sim 10^{-10}$. Maximum error across $18$ output entries is $1.5 \times 10^{-10}$. Applying a uniformly random $R \in \mathrm{SO}(3)$ as an active rotation, the output transforms as $\varphi^{\rm rot} = D^{l_{\rm out}}(R)\varphi$ with $p_{\rm out}$ unchanged, to machine precision ($\sim 10^{-15}$).

\subsection{Real-basis $\kappa^{-1}$ extraction and basis consistency}
\label{sec:numerical-realsh}

We repeat the $\kappa^{-1}$-extraction in the real basis Eq.\eqref{eq:realY}, with $G_{\rm real}$ and $Q_{\rm real}$ defined directly by integration with real spherical harmonics. Both are real-valued at every entry, so $\kappa^{-1}_{\rm real}(l_1, l_2, l) = Q_{\rm real}/G_{\rm real}$ is real with magnitude unitary-invariant, $|\kappa^{-1}_{\rm real}(l_1, l_2, l)| = |\kappa^{-1}_{\rm complex}(l_1, l_2, l)|$. Across all parity-odd triples from $(1,1,1)$ to $(4,4,7)$, magnitudes match across bases to within $1\times 10^{-10}$. Representative values are $|\kappa^{-1}(1,1,1)| = \sqrt{3/(2\pi)} \approx 0.6910$ (analytic value Eq.\eqref{eq:kappa-inv-111}), $|\kappa^{-1}(2,1,2)| \approx 1.1968$, $|\kappa^{-1}(2,2,1)| \approx 1.5451$, $|\kappa^{-1}(3,3,1)| \approx 2.5854$, $|\kappa^{-1}(4,4,7)| \approx 6.2662$. The real-basis pipeline of Appendix~\ref{sec:realsh} runs in real arithmetic end to end.

\section{Real spherical harmonics}
\label{sec:realsh}

The pipelines in Secs.~\ref{sec:pipeline-lebedev}, \ref{sec:pipeline-structured}, and~\ref{sec:pipeline-msgpass} were stated in the complex SH basis. When the input fields $A^{(s)}_{inlm}$ are stored in a real SH basis, the same pipelines run unchanged in algebraic structure with the complex Fourier transform replaced by its real cosine/sine counterpart, giving a $\sim 2\times$ speed-up on the azimuthal stages.

The real spherical harmonics are
\begin{equation}
\label{eq:realY}
Y_{lm}(\theta, \phi) = 
\begin{cases} 
\sqrt{2}\,\tilde{P}_l^{|m|}(\cos\theta)\,\cos(m\phi) & m > 0 \\
\tilde{P}_l^{0}(\cos\theta) & m = 0 \\
\sqrt{2}\,\tilde{P}_l^{|m|}(\cos\theta)\,\sin(|m|\phi) & m < 0 
\end{cases}
\end{equation}
The polar component $p_{ulm} = \tilde{P}_l^{|m|}(\cos\theta_u)$ is unchanged. The azimuthal component is
\begin{equation}
\label{eq:realq}
q_{vm} = 
\begin{cases} 
\sqrt{2}\,\cos(m\phi_v) & m > 0 \\
1 & m = 0 \\
\sqrt{2}\,\sin(|m|\phi_v) & m < 0 
\end{cases}
\end{equation}
which is real-valued ($q_{vm}^* = q_{vm}$). The Lebedev SH values $y_{klm}$ are likewise real.

With these substitutions, we have
\begin{enumerate}
\item In the Lebedev pipeline, $\tilde w_{klm} = w_k\,y_{klm}^*$ reduces to $\tilde w_{klm} = w_k\,y_{klm}$ and all tensors are real.
\item In the structured-grid pipeline, the forward and inverse Fourier transforms in Eqs.\eqref{eq:struct-tildeA} and~\eqref{eq:struct-Tmu} become real cosine/sine transforms. The rest is unchanged.
\item In the parity-aware pipeline, the same applies. The surface-curl construction Eq.\eqref{eq:cat-beta-curl} uses only pointwise vector operations and is basis-independent, with $\partial_\phi$ acting on $q_{vm}$ as a real derivative ($\sqrt{2}\cos(m\phi) \leftrightarrow -m\sqrt{2}\sin(m\phi)$). The $m \leftrightarrow -m$ branching of the real basis requires care in the implementation. $\partial_\phi$ maps the $m > 0$ cosine slot to the $m < 0$ sine slot (and vice versa with a sign), and the gradient cross product mixes the two branches.
\end{enumerate}
Real cosine/sine transforms have roughly half the memory footprint and double the throughput of complex FFTs, accelerating the azimuthal stages by $\sim 2\times$. The remaining stages (radial contractions, polar Legendre transforms, pointwise products, surface-curl pairings, final channel projection) are unchanged in structure and cost. The asymptotic $\mathcal{O}(L^3)$ scaling is set by the radial--angular contractions and the polar Legendre transforms and is preserved.

\section{Per-step wall-clock breakdown of the message-passing pipeline}
\label{sec:profiling-appendix}

To localize the wall-clock cost within the pipeline, we profile each stage of the parity-aware message-passing forward pass (Category $\alpha$, see Sec.~\ref{sec:pipeline-msgpass}) in isolation at $C = 256$, $N_{\rm out} = 4$, $N_{\rm atoms} = 100$, $Z = 50$, fp32, sweeping $l_{\max}$ from $4$ to $22$. The measured log-log scaling slope per stage is given in Table~\ref{tab:profiling-slopes}. The edge and indicator forward transforms denote the combined Legendre--Fourier steps applied to the per-bond edge field [Eqs.~\eqref{eq:msgpass-FE},~\eqref{eq:msgpass-AE}] and the per-atom indicator [Eqs.~\eqref{eq:msgpass-FI},~\eqref{eq:msgpass-AI}]. Indicator gather, pointwise product, and neighbor sum together implement Eq.~\eqref{eq:msgpass-S} (broadcast $j \to ji$, bondwise product, $\sum_j$ reduction). The inverse transform refers to the per-atom inverse Fourier--Legendre steps [Eqs.~\eqref{eq:msgpass-Tmu},~\eqref{eq:msgpass-Tlm}], and channel projection to Eq.~\eqref{eq:cat-alpha-phi}. The per-bond stages with $L^2$ output tensors match their theoretical $L^2$ slope. The per-bond edge forward transform, which carries formally $L^3$ arithmetic, exhibits a measured slope of $1.9$ rather than $3$. At $l_{\max}=22$ the edge forward transform accounts for $19.4$ ms of the $41.5$ ms full forward pass and the neighbor sum contributes a further $16.8$ ms, while the per-atom inverse transform and the channel projection each contribute under $1$ ms. The wall-clock is dominated by the per-bond stages and tracks their measured slopes.
\begin{table}[!htbp]
\centering
\begin{tabular}{lcc}
\hline
Step                       & Theoretical & Measured \\
\hline
Edge forward transform     & $L^3$ & $1.9$ \\
Indicator forward transform & $L^3$ & $0.2$ \\
Indicator gather to bonds  & $L^2$ & $1.6$ \\
Pointwise product          & $L^2$ & $1.6$ \\
Neighbor sum               & $L^2$ & $1.8$ \\
Inverse transform          & $L^3$ & $0.2$ \\
Channel projection         & $L^2$ & $0.0$ \\
\hline
Full forward pass          & $L^3$ & $\mathbf{1.9}$ \\
\hline
\end{tabular}
\caption{Per-step log-log scaling slopes of the parity-aware message-passing pipeline (Category $\alpha$).}
\label{tab:profiling-slopes}
\end{table}

\bibliographystyle{apsrev4-2}
\bibliography{references}

\end{document}